\documentclass[sigplan,screen]{acmart}
\setcopyright{acmlicensed}
\acmPrice{15.00}
\acmDOI{10.1145/3453483.3454033}
\acmYear{2021}
\copyrightyear{2021}
\acmSubmissionID{pldi21main-p43-p}
\acmISBN{978-1-4503-8391-2/21/06}
\acmConference[PLDI '21]{Proceedings of the 42nd ACM SIGPLAN International Conference on Programming Language Design and Implementation}{June 20--25, 2021}{Virtual, Canada}
\acmBooktitle{Proceedings of the 42nd ACM SIGPLAN International Conference on Programming Language Design and Implementation (PLDI '21), June 20--25, 2021, Virtual, Canada}

\usepackage{booktabs}   %% For formal tables:
\usepackage{subcaption} %% For complex figures with subfigures/subcaptions
\usepackage{enumerate} % for lists
\usepackage{listings} % for code
\usepackage{xspace} % so we don't need to figure out spacing after \toolname every time
\usepackage{mathpartir} % for inference rules
\usepackage{dsfont} % to render N for the natural numbers
\usepackage{syntax} % for code highlighting
\usepackage{xcolor} % for code highlighting
\usepackage{colortbl} % to black out table cells
\usepackage{multirow} % for tables
\usepackage{tikz} % for fancy circles and smileys

\lstdefinelanguage{coq}{
    keywords={Repair, module, Theorem, Proof, Record, Lemma, Definition, Abort, Qed, forall, Inductive, Type, Prop, Set, fun, fix, forall, Require, Import, Fixpoint, match, end, with, as, return, struct, Qed, Defined, let},
    basicstyle=\linespread{0.95}\small\ttfamily,
    keywordstyle=\color{blue},
    commentstyle=\itshape\rmfamily,
    showstringspaces=false,
    columns=flexible,
    breaklines=true,
    texcl=true,
    mathescape=true,
    tabsize=4,
    stringstyle=\color{brown},
    escapeinside={(@}{@)},
}

\newcommand\toolname{\textsc{Pumpkin P}i\xspace} % tool name
\newcommand\company{Galois\xspace} % company name
\newcommand\A{$A$\xspace} % recurring type A
\newcommand\B{$B$\xspace} % recurring type B
\newcommand{\reducedstrut}{\vrule width 0pt height .9\ht\strutbox depth .9\dp\strutbox\relax} % for \codediff and \codeauto
\newcommand{\codediff}[1]{%
  \begingroup
  \setlength{\fboxsep}{0pt}%
  \colorbox{orange!25}{\reducedstrut#1\/}%
  \endgroup
} % to highlight the difference between two code blocks
\newcommand{\codesima}[1]{%
  \begingroup
  \setlength{\fboxsep}{0pt}%
  \colorbox{orange!25}{\reducedstrut#1\raisebox{0.8ex}{\scalebox{0.66}{2}}\/}%
  \endgroup
} % to highlight the similarities between two code blocks
\renewcommand{\textsuperscript}[1]{}
\newcommand{\codesimb}[1]{%
  \begingroup
  \setlength{\fboxsep}{0pt}%
  \colorbox{red!25}{\reducedstrut#1\raisebox{0.8ex}{\scalebox{0.66}{1}}\/}%
  \endgroup
} % to highlight the similarities between two code blocks
\newcommand{\codesimc}[1]{%
  \begingroup
  \setlength{\fboxsep}{0pt}%
  \colorbox{yellow!25}{\reducedstrut#1\raisebox{0.8ex}{\scalebox{0.66}{3}}\/}%
  \endgroup
} % to highlight the similarities between two code blocks
\newcommand{\codesimd}[1]{%
  \begingroup
  \setlength{\fboxsep}{0pt}%
  \colorbox{green!25}{\reducedstrut#1\raisebox{0.8ex}{\scalebox{0.66}{4}}\/}%
  \endgroup
} % to highlight the similarities between two code blocks
\newcommand{\codesime}[1]{%
  \begingroup
  \setlength{\fboxsep}{0pt}%
  \colorbox{blue!25}{\reducedstrut#1\raisebox{0.8ex}{\scalebox{0.66}{5}}\/}%
  \endgroup
} % to highlight the similarities between two code blocks
\newcommand{\codeauto}[1]{%
  \begingroup
  \setlength{\fboxsep}{0pt}%
  \colorbox{cyan!30}{\reducedstrut#1\/}%
  \endgroup
} % to highlight automatically-generated terms
\newcommand{\mysubsubsec}[1]{\vspace{0.40em} \noindent {{\textbf{#1.}}}}

\newcommand*\circled[1]{\tikz[baseline=(char.base)]{
            \node[shape=circle,draw,inner sep=0.5pt] (char) {#1};}}

\DeclareRobustCommand\good{\tikz[baseline=(char.base)]{
    \draw circle (1.6mm);
\node[fill,circle,inner sep=0.5pt] (left eye) at (135:0.8mm) {};
\node[fill,circle,inner sep=0.5pt] (right eye) at (45:0.8mm) {};
\draw (-145:0.8mm) arc (-145:-35:0.8mm);
}}

\DeclareRobustCommand\ok{\tikz[baseline=(char.base)]{
    \draw circle (1.6mm);
\node[fill,circle,inner sep=0.5pt] (left eye) at (135:0.8mm) {};
\node[fill,circle,inner sep=0.5pt] (right eye) at (45:0.8mm) {};
\draw (-135:0.9mm) -- (-45:0.9mm);
}}

\DeclareRobustCommand\bad{\tikz[baseline=(char.base)]{
    \draw circle (1.6mm);
\node[fill,circle,inner sep=0.5pt] (left eye) at (135:0.8mm) {};
\node[fill,circle,inner sep=0.5pt] (right eye) at (45:0.8mm) {};
\draw (-135:0.9mm) arc (145:35:0.8mm);
}}

\begin{document}

%% Title information
\title{Proof Repair across Type Equivalences}         %% [Short Title] is optional;
                                        %% when present, will be used in
                                        %% header instead of Full Title.
%\titlenote{}             %% \titlenote is optional;
                                        %% can be repeated if necessary;
                                        %% contents suppressed with 'anonymous'
%\subtitle{Subtitle}                     %% \subtitle is optional
%\subtitlenote{with subtitle note}       %% \subtitlenote is optional;
                                        %% can be repeated if necessary;
                                        %% contents suppressed with 'anonymous'

%% Author information
%% Contents and number of authors suppressed with 'anonymous'.
%% Each author should be introduced by \author, followed by
%% \authornote (optional), \orcid (optional), \affiliation, and
%% \email.
%% An author may have multiple affiliations and/or emails; repeat the
%% appropriate command.
%% Many elements are not rendered, but should be provided for metadata
%% extraction tools.

%% Author with single affiliation.
\author{Talia Ringer}

\affiliation{
  \institution{University of Washington}  
  \country{USA}                    %% \country is recommended
}
\email{tringer@cs.washington.edu}          %% \email is recommended

%% Author with two affiliations and emails.
\author{RanDair Porter}

\affiliation{
  \institution{University of Washington}  
  \country{USA}                    %% \country is recommended
}
\email{randair@uw.edu}         %% \email is recommended

\author{Nathaniel Yazdani}

\affiliation{
  \institution{Northeastern University}  
  \country{USA}                    %% \country is recommended
}
\email{yazdani.n@husky.neu.edu}         %% \email is recommended

\author{John Leo}

\affiliation{
  \institution{Halfaya Research}  
  \country{USA}                    %% \country is recommended
}
\email{leo@halfaya.org}         %% \email is recommended

\author{Dan Grossman}

\affiliation{
  \institution{University of Washington}  
  \country{USA}                    %% \country is recommended
}
\email{djg@cs.washington.edu}         %% \email is recommended

%% Abstract
%% Note: \begin{abstract}...\end{abstract} environment must come
%% before \maketitle command
\begin{abstract}
We describe a new approach to automatically repairing broken proofs in the Coq proof assistant in response to changes in types.
Our approach combines a configurable proof term transformation with a decompiler from proof terms to suggested tactic scripts.
The proof term transformation implements transport across equivalences in a way that removes references to the old version of the changed type and does not rely on axioms beyond those Coq assumes.

We have implemented this approach in \toolname, an extension to the \textsc{Pumpkin Patch} Coq plugin suite for proof repair.
We demonstrate \toolname's flexibility on eight case studies,
including supporting a benchmark from a user study,
easing development with dependent types,
porting functions and proofs between unary and binary numbers,
and supporting an industrial proof engineer to interoperate between Coq and other verification tools more easily.
\end{abstract}



%% Keywords
%% comma separated list
  %% \keywords are mandatory in final camera-ready submission

%% \maketitle
%% Note: \maketitle command must come after title commands, author
%% commands, abstract environment, Computing Classification System
%% environment and commands, and keywords command.
\maketitle

%% Body
\section{Introduction}

Program verification with interactive theorem provers has come a long way since its inception,
especially when it comes to the scale of programs that can be verified.
The seL4~\cite{Klein2009} verified operating system kernel, for example,
is the effort of a team of proof engineers spanning more than
a million lines of proof, costing over 20 person-years.
Given a famous 1977 critique of verification~\cite{DeMillo1977} (emphasis ours):

\begin{quote}
\textit{A sufficiently fanatical researcher}
might be willing to devote \textit{two or 
three years} to verifying a significant 
piece of software if he could be 
assured that the software would remain stable.
\end{quote}
we could argue that, over 40 years, either verification has become easier,
or researchers have become more fanatical. Unfortunately, not all has changed (emphasis still ours):

\begin{quote}
But real-life programs need to 
be maintained and modified. 
There is \textit{no reason to believe} that verifying a modified program is any 
easier than verifying the original the 
first time around.
\end{quote}
Tools that can automatically refactor or repair proofs~\cite{wibergh2019, WhitesidePhD, Dietrich2013, adams2015, Bourke12, Roe2016, robert2018, pumpkinpatch}
give us reason to believe that verifying a modified program \textit{can} sometimes be easier than verifying the original, even when proof engineers do not follow good development processes,
or when change occurs outside of proof engineers' control~\cite{PGL-045}.
Still, maintaining verified programs can be challenging: it means keeping not just the programs, but also specifications and proofs about those programs up-to-date.
This remains so difficult that sometimes, even experts give up in the face of change~\cite{replica}.

The problem of automatically updating proofs in response to changes in programs or specifications is known as \textit{proof repair}~\cite{PGL-045, pumpkinpatch}.
While there are many ways proofs need to be repaired, one such need is in response to a changed type definition (Section~\ref{sec:key1}).
We make progress on two open challenges in proof repair in response to changes in type definitions:

\begin{enumerate}
\item Existing work supports very limited classes of these changes like non-structural changes~\cite{pumpkinpatch} or a predefined set
of changes~\cite{robert2018, wibergh2019}, and these are not informed by the needs of proof engineers~\cite{replica}.
\item Proof repair tools are not yet integrated with typical proof engineering workflows~\cite{PGL-045, pumpkinpatch, robert2018},
and may impose additional proof obligations like proving relations corresponding to changes~\cite{Ringer2019}.
\end{enumerate}

The typical proof engineering workflow in Coq is interactive:
The proof engineer passes Coq high-level search procedures called \textit{tactics} (like \lstinline{induction}),
and Coq responds to each tactic by refining the current goal to some subgoal (like the proof obligation for the base case).
This loop of tactics and goals continues until no goals remain, at which point the proof engineer 
has constructed a sequence of tactics called a \textit{proof script}.
To check this proof script for correctness, Coq compiles it to a low-level representation called a \textit{proof term},
then checks that the proof term has the expected type.

Our approach to proof repair works at the level of low-level proof terms, then builds back up to high-level proof scripts.
In particular, our approach combines a configurable proof term transformation (Section~\ref{sec:key2}) with a prototype decompiler from proof terms
back to suggested proof scripts (Section~\ref{sec:decompiler}). % TODO mention search or not?
This is implemented (Section~\ref{sec:impl}) in \toolname, an extension to the \textsc{Pumpkin Patch}~\cite{pumpkinpatch} proof repair plugin suite for Coq 8.8 that is available on Github.\footnote{We annotate each claim to which code is relevant with a circled number like \circled{1}. These circled numbers are links to code, and are detailed in \href{https://github.com/uwplse/pumpkin-pi/blob/v2.0.0/GUIDE.md}{\lstinline{GUIDE.md}}.} % TODO links here
%The result is a flexible proof repair tool that: 

%\begin{enumerate}
%\item supports changes in types informed by proof engineers and not supported by other tools, and
%\item suggests tactic scripts and proves relations corresponding to certain changes for better workflow integration.
%\end{enumerate}

\mysubsubsec{Addressing Challenge 1: Flexible Type Support}
The case studies in Section~\ref{sec:search}---summarized in Table~\ref{fig:changes} on page~\pageref{fig:changes}---show that \toolname is flexible enough to support
a wide range of proof engineering use cases. % can support a flexible class of changes informed by the needs of proof engineers within a unified framework.
In general, \toolname can support any change described by an equivalence, though it takes the equivalence in a
deconstructed form that we call a \textit{configuration}.
The configuration expresses to the proof term transformation how to translate functions and proofs defined over the old version of a type
to refer only to the new version, and how to do so in a way that does not break definitional equality.
The proof engineer can write this configuration in Coq and feed it to \toolname (\textit{manual configuration} in Table~\ref{fig:changes}),
configuring \toolname to support the change. %from directly within Coq.

\mysubsubsec{Addressing Challenge 2: Workflow Integration}
Research on workflow integration for proof repair tools is in its infancy.
\toolname is built with workflow integration in mind.
For example, \toolname is the only proof repair tool we are aware of that produces suggested proof scripts (rather than proof terms) for repaired proofs,
a challenge highlighted in existing proof repair work~\cite{pumpkinpatch, robert2018} and in 
a survey of proof engineering~\cite{PGL-045}.
In addition, \toolname implements search procedures that 
automatically discover configurations and prove the equivalences they induce for four different classes of 
changes (\textit{automatic configuration} in Table~\ref{fig:changes}),
decreasing the burden of proof obligations imposed on the proof engineer.
Our partnership with an industrial proof engineer has informed other changes to further improve workflow integration
(Sections~\ref{sec:implementation} and~\ref{sec:search}).

\begin{figure}
\includegraphics[width=\columnwidth]{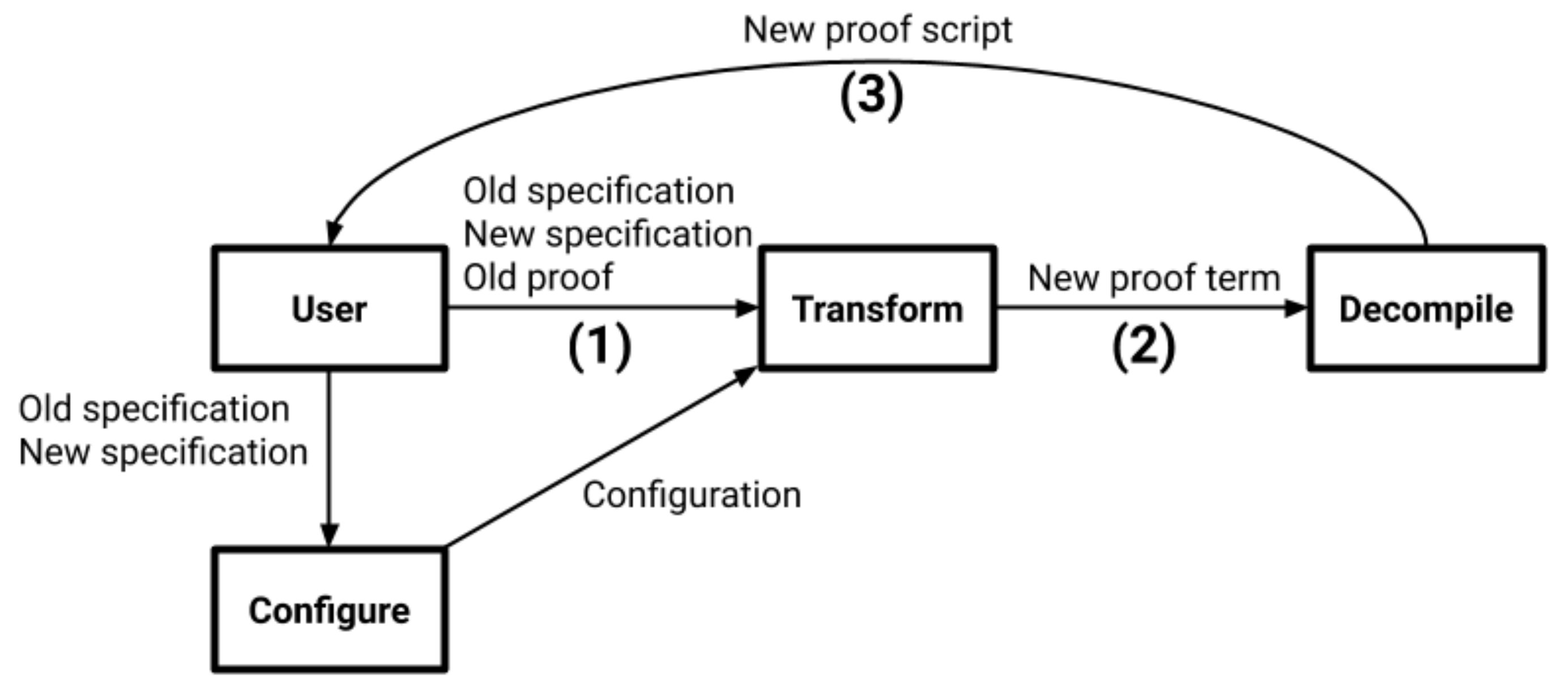}
\vspace{-0.7cm}
\caption{The workflow for \toolname.}
\vspace{-0.1cm}
\label{fig:system}
\end{figure}

\mysubsubsec{Bringing it Together}
Figure~\ref{fig:system} shows how this comes together when the proof engineer invokes \toolname:

\begin{enumerate}
\item The proof engineer \textbf{Configure}s \toolname, either manually or automatically.
\item The configured \textbf{Transform} transforms the old proof term into the new proof term.
\item \textbf{Decompile} suggests a new proof script. % given the new proof term.
\end{enumerate}
There are currently four search procedures for automatic configuration implemented in \toolname (see Table~\ref{fig:changes} on page~\pageref{fig:changes}).
%All four search procedures generate equivalence proofs as in Figure~\ref{fig:equivalence} automatically (\href{https://github.com/uwplse/pumpkin-pi/%blob/master/plugin/src/automation/search/search.ml}{search.ml} and \href{https://github.com/uwplse/pumpkin-pi/blob/master/plugin/src/automation/search/equivalence.ml}{equivalence.ml}),
%then configure (\href{https://github.com/uwplse/pumpkin-pi/blob/master/plugin/src/automation/lift/liftconfig.ml}{liftconfig.ml}) the transformation %to those equivalences.
Manual configuration makes it possible
for the proof engineer to configure the transformation to any equivalence,
even without a search procedure.
Section~\ref{sec:search} shows examples of both workflows applied to real scenarios.

%Our main technical advances are techniques for transforming proof terms directly to repair broken proofs in response to changes
%in types, while our prototype decompiler up to tactics is important for usability in Coq.
%We elaborate on some of the examples from Table~\ref{fig:changes} to further demonstrate flexibility and usability with four case studies (Section~%\ref{sec:search}), which show that \toolname 1) can support a benchmark from a user study of Coq proof engineers, 2) can simplify dependently-typed programming, %automating manual steps from previous work,
%3) can help proof engineers port functions and proofs from unary to binary numbers, and
%4) has helped an industrial proof engineer at Galois integrate Coq with a company workflow and write proofs about an implementation of the TLS %Handshake Protocol.

\section{A Simple Motivating Example}
\label{sec:overview}

\begin{figure*}
\begin{minipage}{0.46\textwidth}
\begin{lstlisting}[firstline=1, lastline=3]
Inductive list (T : Type) : Type :=(@\vspace{-0.04cm}@)
| (@\codediff{nil}@) : list T(@\vspace{-0.04cm}@)
| (@\codediff{cons}@) :  T $\rightarrow$ list T $\rightarrow$ list T.(@\vspace{-0.04cm}@)
\end{lstlisting}
\end{minipage}
\hfill
\begin{minipage}{0.46\textwidth}
\begin{lstlisting}[firstline=5, lastline=7]
Inductive list (T : Type) : Type :=(@\vspace{-0.04cm}@)
| (@\codediff{cons}@) : T $\rightarrow$ list T $\rightarrow$ list T(@\vspace{-0.04cm}@)
| (@\codediff{nil}@) : list T.(@\vspace{-0.04cm}@)
\end{lstlisting}
\end{minipage}
\vspace{-0.4cm}
\caption{A change from the old version of \lstinline{list} (left) to the new version of \lstinline{list} (right).
The old version of \lstinline{list} is an inductive datatype that is either empty (the \lstinline{nil} constructor), or the result
of placing an element in front of another \lstinline{list} (the \lstinline{cons} constructor). The change swaps these
two constructors (\codediff{orange}).}
\label{fig:listswap}
\end{figure*}

\begin{figure*}
\codeauto{%
\begin{minipage}{0.48\textwidth}
\begin{lstlisting}[firstline=1, lastline=13]
swap$\phantom{^{2}}$T (l : Old.list T) : New.list T :=(@\vspace{-0.04cm}@)
  Old.list_rect T (fun (l : Old.list T) => New.list T)(@\vspace{-0.04cm}@)
    New.nil(@\vspace{-0.04cm}@)
    (fun t _ (IHl : New.list T) => New.cons T t IHl)(@\vspace{-0.04cm}@)
    l.(@\vspace{-0.04cm}@)
(@\vspace{-0.14cm}@)
Lemma section: $\forall$ T (l : Old.list T),(@\vspace{-0.04cm}@)
  swap$^{-1}$ T (swap T l) = l.(@\vspace{-0.04cm}@)
Proof.(@\vspace{-0.04cm}@)
  intros T l. symmetry. induction l as [ |a l0 H].(@\vspace{-0.04cm}@)
  - auto.(@\vspace{-0.04cm}@)
  - simpl. rewrite $\leftarrow$ H. auto.(@\vspace{-0.04cm}@)
Qed.(@\vspace{-0.04cm}@)
\end{lstlisting}
\end{minipage}
\hfill
\begin{minipage}{0.48\textwidth}
\begin{lstlisting}[firstline=15, lastline=28]
swap$^{-1}$ T (l : New.list T) : Old.list T :=(@\vspace{-0.04cm}@)
  New.list_rect T (fun (l : New.list T) => Old.list T)(@\vspace{-0.04cm}@)
    (fun t _ (IHl : Old.list T) => Old.cons T t IHl)(@\vspace{-0.04cm}@)
    Old.nil(@\vspace{-0.04cm}@)
    l.(@\vspace{-0.04cm}@)
(@\vspace{-0.14cm}@)
Lemma retraction: $\forall$ T (l : New.list T),(@\vspace{-0.04cm}@)
  swap T (swap$^{-1}$ T l) = l.(@\vspace{-0.04cm}@)
Proof.(@\vspace{-0.04cm}@)
  intros T l. symmetry. induction l as [t l0 H| ].(@\vspace{-0.04cm}@)
  - simpl. rewrite $\leftarrow$ H. auto.(@\vspace{-0.04cm}@)
  - auto.(@\vspace{-0.04cm}@)
Qed.(@\vspace{-0.04cm}@)
\end{lstlisting}
\end{minipage}}
\vspace{-0.3cm}
\caption{Two functions between \lstinline{Old.list} and \lstinline{New.list} (top) that form an equivalence (bottom).}
\label{fig:equivalence}
\end{figure*}

Consider a simple example of using \toolname: repairing proofs after swapping the two constructors of the \lstinline{list} datatype (Figure~\ref{fig:listswap}).
This is inspired by a similar change from a user study of proof engineers (Section~\ref{sec:search}).
Even such a simple change can cause trouble, as in this proof from the Coq standard library (comments ours for clarity):\footnote{We use induction instead of pattern matching.}

\begin{lstlisting}
Lemma rev_app_distr {A} :(@\vspace{-0.04cm}@)
  $\forall$ (x y : list A), rev (x ++ y) = rev y ++ rev x.(@\vspace{-0.04cm}@)
Proof. (@\color{gray}{(* by induction over x and y *)}@)(@\vspace{-0.04cm}@)
  induction x as [| a l IHl].(@\vspace{-0.04cm}@)
  (@\color{gray}{(* x nil: *)}@) induction y as [| a l IHl].(@\vspace{-0.04cm}@)
  (@\color{gray}{(* y nil: *)}@) simpl. auto.(@\vspace{-0.04cm}@)
  (@\color{gray}{(* y cons *)}@) simpl. rewrite app_nil_r; auto.(@\vspace{-0.04cm}@)
  (@\color{gray}{(* both cons: *)}@) intro y. simpl.(@\vspace{-0.04cm}@)
  rewrite (IHl y). rewrite app_assoc; trivial.(@\vspace{-0.04cm}@)
Qed.(@\vspace{-0.05cm}@)
\end{lstlisting}
This lemma says that appending (\lstinline{++}) two lists and reversing (\lstinline{rev}) the result behaves the same as appending
the reverse of the second list onto the reverse of the first list.
The proof script works by induction over the input lists \lstinline{x} and \lstinline{y}:
In the base case for both \lstinline{x} and \lstinline{y}, the result holds by reflexivity.
In the base case for \lstinline{x} and the inductive case for \lstinline{y}, the result follows from the existing lemma \lstinline{app_nil_r}.
Finally, in the inductive case for both \lstinline{x} and \lstinline{y}, the result follows by the inductive hypothesis
and the existing lemma \lstinline{app_assoc}.

When we change the \lstinline{list} type, this proof no longer works.
%This theorem statement \lstinline{rev_app_distr} defined over the old version of \lstinline{list} is our \textit{old specification}.
%When we change the \lstinline{list} type, we get the \textit{new specification}.
%But the \textit{old proof} or tactic script no longer works with this new specification.
To repair this proof with \toolname, we run this command:

\iffalse
\begin{figure}
\begin{minipage}{0.46\textwidth}
\begin{lstlisting}[firstline=1, lastline=3]
Inductive list (T : Type) : Type :=(@\vspace{-0.04cm}@)
| (@\codediff{nil}@) : list T(@\vspace{-0.04cm}@)
| (@\codediff{cons}@) :  T $\rightarrow$ list T $\rightarrow$ list T.(@\vspace{-0.04cm}@)
\end{lstlisting}
\end{minipage}
\hfill
\begin{minipage}{0.46\textwidth}
\begin{lstlisting}[firstline=5, lastline=7]
Inductive list (T : Type) : Type :=(@\vspace{-0.04cm}@)
| (@\codediff{cons}@) : T $\rightarrow$ list T $\rightarrow$ list T(@\vspace{-0.04cm}@)
| (@\codediff{nil}@) : list T.(@\vspace{-0.04cm}@)
\end{lstlisting}
\end{minipage}
\vspace{-0.4cm}
\caption{The updated \lstinline{list} (bottom) is the old \lstinline{list} (top) with its two constructors swapped (\codediff{orange}).}
\label{fig:listswap}
\end{figure}
\fi

\begin{lstlisting}
Repair Old.list New.list in rev_app_distr.(@\vspace{-0.05cm}@)
\end{lstlisting}
assuming the old and new list types from Figure~\ref{fig:listswap} are in modules \lstinline{Old} and \lstinline{New}.
This suggests a proof script that succeeds (in $\codeauto{light blue}$ to denote \toolname produces it automatically):

\begin{lstlisting}[backgroundcolor=\color{cyan!30}]
Proof. (@\color{gray}{(* by induction over x and y *)}@)(@\vspace{-0.04cm}@)
  intros x. induction x as [a l IHl| ]; intro y0.(@\vspace{-0.04cm}@)
  - (@\color{gray}{(* both cons: *)}@) simpl. rewrite IHl. simpl.(@\vspace{-0.04cm}@)
    rewrite app_assoc. auto.(@\vspace{-0.04cm}@)
  - (@\color{gray}{(* x nil: *)}@) induction y0 as [a l H| ].(@\vspace{-0.04cm}@)
    + (@\color{gray}{(* y cons: *)}@) simpl. rewrite app_nil_r. auto.(@\vspace{-0.04cm}@)
    + (@\color{gray}{(* y nil: *)}@) auto.(@\vspace{-0.04cm}@)
Qed.(@\vspace{-0.05cm}@)
\end{lstlisting}
where the dependencies (\lstinline{rev}, \lstinline{++}, \lstinline{app_assoc}, and \lstinline{app_nil_r}) have
also been updated automatically~\href{https://github.com/uwplse/pumpkin-pi/blob/v2.0.0/plugin/coq/Swap.v}{\circled{1}}. % Swap.v
If we would like, we can manually modify this to something that more closely matches the style of the original proof script:

\begin{lstlisting}
Proof. (@\color{gray}{(* by induction over x and y *)}@)(@\vspace{-0.04cm}@)
  induction x as [a l IHl|].(@\vspace{-0.04cm}@)
  (@\color{gray}{(* both cons: *)}@) intro y. simpl.(@\vspace{-0.04cm}@)
  rewrite (IHl y). rewrite app_assoc; trivial.(@\vspace{-0.04cm}@)
  (@\color{gray}{(* x nil: *)}@) induction y as [a l IHl|].(@\vspace{-0.04cm}@)
  (@\color{gray}{(* y cons: *)}@) simpl. rewrite app_nil_r; auto.(@\vspace{-0.04cm}@)
  (@\color{gray}{(* y nil: *)}@) simpl. auto.(@\vspace{-0.04cm}@)
Qed.(@\vspace{-0.04cm}@)
\end{lstlisting}
We can even repair the entire list module from the Coq standard library all at once by running the \lstinline{Repair module}
command~\href{https://github.com/uwplse/pumpkin-pi/blob/v2.0.0/plugin/coq/Swap.v}{\circled{1}}. % Swap.v
When we are done, we can get rid of \lstinline{Old.list}. % entirely.

The key to success is taking advantage of Coq's structured proof term language:
Coq compiles every proof script to a proof term in a rich functional programming language called 
%Gallina that is based on the calculus of inductive  constructions---\toolname repairs that term.
Gallina---\toolname repairs that term.
\toolname then decompiles the repaired proof term (with optional hints from the original proof script) back 
to a suggested proof script that the proof engineer can maintain.
%Here, \toolname transforms the proof term Coq compiles \lstinline{rev_app_distr} to,
%and then decompiles that transformed proof term to the proof script in light blue.

In contrast, updating the poorly structured proof script directly would not be straightforward.
Even for the simple proof script above, grouping tactics by line, there are $6! = 720$ permutations of this proof script.
It is not clear which lines to swap since these tactics do not have a semantics beyond the searches their evaluation performs.
Furthermore, just swapping lines is not enough: even for such a simple change, we must also swap
arguments, so \lstinline{induction x as [| a l IHl]} becomes \lstinline{induction x as [a l IHl|]}.
%Handling even swapping constructors this way would require a search procedure that would not generalize to other changes.
\citet{robert2018} describes the challenges of repairing tactics in detail.
\toolname's approach circumvents this challenge.

%\toolname's approach circumvents this challenge. % by transforming proof terms, then decompiling the transformed proof term to a tactic script.
%By decompiling the transformed proof term, \toolname is able to suggest a tactic script in the end.
%As later sections show, this approach is much more general than just permuting constructors.

\section{Problem Definition}
\label{sec:key1}

\toolname can do much more than permute constructors. % approach is more general than just permuting constructors:
Given an equivalence between types \A and \B,
\toolname repairs functions and proofs defined over \A to instead refer to \B (Section~\ref{sec:scope}).
It does this in a way that allows for removing references to \A, which is essential for proof repair,
since \A may be an old version of an updated type (Section~\ref{sec:repair}).

 %We can view proof repair as a form of 
%\textit{proof reuse}~\cite{Ringer2019, felty1994generalization, caplan1995logical, pons2000generalization, johnsen2004theorem}, % TODO consider citation list
%or reusing proofs about one specification (say, from another library, or from within the same proof development)
%to derive proofs about another specification.
%The difference is that in standard proof reuse, both of these specifications continue to exist.
%In contrast, proof repair is the process of reusing proofs across \textit{two versions of a single specification},
%only one of which---the new version---must continue to exist.
%That is, the old version of the specification may be removed after updating proofs to use the new version (Section~\ref{sec:repair}).
%The key to supporting proof repair is to build a proof reuse tool that can handle that additional challenge (Section~\ref{sec:time}).

%\begin{quote}
%\textbf{Insight 1}:
%Proof repair is a form of proof reuse---reusing proofs about one specification to derive proofs about another specification---with 
%the additional challenge that one of the specifications may cease to exist (Section~\ref{sec:repair}).
%The key to supporting proof repair is to build a proof reuse
%tool that can handle that additional challenge (Section~\ref{sec:time}).
%\end{quote}

\subsection{Scope: Type Equivalences}
\label{sec:scope}

\toolname repairs proofs in response to changes in types that correspond to \textit{type equivalences}~\cite{univalent2013homotopy},
or pairs of functions that map between two types and are mutual inverses.\footnote{The adjoint follows, and \toolname includes machinery to prove it~\href{https://github.com/uwplse/pumpkin-pi/blob/v2.0.0/plugin/src/automation/search/equivalence.ml}{\circled{10}}~\href{https://github.com/uwplse/pumpkin-pi/blob/v2.0.0/plugin/theories/Adjoint.v}{\circled{23}}.}
When a type equivalence between types \A and \B exists, those types are \textit{equivalent} (denoted \A $\simeq$ \B). % for example:
Figure~\ref{fig:equivalence} shows a type equivalence between the two versions of \lstinline{list}
from Figure~\ref{fig:listswap} that \toolname discovered and proved automatically~\href{https://github.com/uwplse/pumpkin-pi/blob/v2.0.0/plugin/coq/Swap.v}{\circled{1}}.

%
%\begin{lstlisting}
%Old.list $\simeq$ New.list
%\end{lstlisting}

To give some intuition for what kinds of changes can be described by equivalences, we preview two changes below.
See Table~\ref{fig:changes} on page~\pageref{fig:changes} for more examples.

\mysubsubsec{Factoring out Constructors}
Consider changing the type \lstinline{I} to the type \lstinline{J} 
in Figure~\ref{fig:equivalence2}.
\lstinline{J} can be viewed as \lstinline{I} with its two constructors \lstinline{A} and \lstinline{B} pulled out to a
new argument of type \lstinline{bool} for a single constructor.
With \toolname, the proof engineer can repair functions and proofs about \lstinline{I} to instead use \lstinline{J},
as long as she configures \toolname to describe which constructor 
of \lstinline{I} maps to \lstinline{true} and which maps to \lstinline{false}.
This information about constructor mappings induces an equivalence \lstinline{I }$\simeq$\lstinline{ J}
across which \toolname repairs functions and proofs.
File \href{https://github.com/uwplse/pumpkin-pi/blob/v2.0.0/plugin/coq/playground/constr_refactor.v}{\circled{2}} shows an example of this, mapping \lstinline{A} to \lstinline{true} and \lstinline{B} to false,
and repairing proofs of De Morgan's laws. % constr_refactor.v
%
%It uses \toolname to automatically repair functions and proofs over \lstinline{I}, like:

%\begin{lstlisting}
%Theorem demorgan_1 : $\forall$ (i1 i2 : I),(@\vspace{-0.04cm}@)
%  neg (and i1 i2) = or (neg i1) (neg i2).(@\vspace{-0.04cm}@)
%Proof.(@\vspace{-0.04cm}@)
%  intros i1 i2.(@\vspace{-0.04cm}@)
%  induction i1; auto.(@\vspace{-0.04cm}@)
%Qed.
%\end{lstlisting}
%to corresponding functions and proofs over \lstinline{J}, like:
%
%\begin{lstlisting}[backgroundcolor=\color{cyan!30}]
%Theorem demorgan_1 : $\forall$ (j1 j2 : J),(@\vspace{-0.04cm}@)
%  neg (and j1 j2) = or (neg j1) (neg j2).(@\vspace{-0.04cm}@)
%Proof.(@\vspace{-0.04cm}@)
%  intros j1 j2.(@\vspace{-0.04cm}@)
%  induction j1 (@\codediff{as [b]. induction b as [ | ]}@); auto.(@\vspace{-0.04cm}@)
%Qed.
%\end{lstlisting}
%These repaired functions and proofs refer to \lstinline{J} in place of \lstinline{I}.
%Otherwise, they behave the same way as the functions and proofs over \lstinline{I} up to the equivalence between
%\lstinline{I} and \lstinline{J}---Section~\ref{sec:repair} explains this intuition more formally.

\mysubsubsec{Adding a Dependent Index}
At first glance, the word \textit{equivalence} may seem to imply that \toolname can support only changes in
which the proof engineer does not add or remove information.
But equivalences are more powerful than they may seem.
%The idea is, when possible, to separate out the new information
%into a projection of a $\Sigma$ type or a constructor of a sum type.
%roofs about this new information become the proof obligation for the proof engineer,
%and \toolname automates the rest.
Consider, for example, changing a list to a length-indexed vector (Figure~\ref{fig:listtovect}).
\toolname can repair functions and proofs about lists to functions and proofs about vectors of particular lengths~\href{https://github.com/uwplse/pumpkin-pi/blob/v2.0.0/plugin/coq/examples/Example.v}{\circled{3}}, % Example.v
since $\Sigma$\lstinline{(l:list T).length l = n }$\simeq$\lstinline{ vector T n}.
From the proof engineer's perspective, after updating specifications from \lstinline{list} to \lstinline{vector},
to fix her functions and proofs, she must additionally prove invariants about the lengths of her lists.
\toolname makes it easy to separate out that proof obligation, then automates the rest.

More generally, in homotopy type theory, with the help of quotient types, it is possible to form an equivalence
from a relation, even when the relation is not an equivalence~\cite{angiuli2020internalizing}.
While Coq lacks quotient types,
it is possible to achieve a similar outcome and use \toolname for changes that add or remove information
when those changes can be expressed as equivalences between $\Sigma$ types or sum types.
%
%(\href{https://github.com/uwplse/pumpkin-pi/blob/master/plugin/coq/examples/Example.v}{\lstinline{Example.v}}).%%
%
%With the proof reuse tool \textsc{Devoid}~\cite{Ringer2019},
%it is possible to repair proofs about lists to proofs about vectors of \textit{some} length, since:
%
%\begin{lstlisting}
%packed_vect T := $\Sigma$(n : nat).vector T n.
%list T $\simeq$ packed_vector T.
%\end{lstlisting}
%This is enough to automatically repair a lemma about lists:
%
%\begin{lstlisting}
%$\forall$ {A B} (l1 : list A) (l2 : list B),(@\vspace{-0.04cm}@)
%  zip_with pair l1 l2 = zip l1 l2.
%\end{lstlisting}
%to a lemma about vectors of some length:
%
%\begin{lstlisting}
%$\forall$ {A B} (l1 : (@\codediff{packed\_vect A}@)) (l2 : (@\codediff{packed\_vect B}@)),(@\vspace{-0.04cm}@)
%  zip_with pair l1 l2 = zip l1 l2.
%\end{lstlisting}
%recursively updating dependencies \lstinline{zip} and \lstinline{zip_with}.
%It is not enough, however, to help the proof engineer get from that to a proof about vectors \textit{of a particular length}:
%
%\begin{lstlisting}
%$\forall$ {A B} (@\codediff{n}@) (l1 : (@\codediff{vector A n}@)) (l2 : (@\codediff{vector B n}@)),(@\vspace{-0.04cm}@)
%  zip_with pair (@\codediff{n}@) l1 l2 = zip (@\codediff{n}@) l1 l2.
%\end{lstlisting}
%\textsc{Devoid} leaves this step to the proof engineer.
%\toolname, in contrast, can handle this step as well (\href{https://github.com/uwplse/pumpkin-pi/blob/master/plugin/coq/examples/Example.v}{\lstinline{Example.v}}).
%The key is to repair functions and proofs across this equivalence:

%Section~\ref{sec:search} shows this and other case studies using \toolname to repair real proofs
%informed by the needs of proof engineers.

\subsection{Goal: Transport with a Twist}
\label{sec:repair}

The goal of \toolname is to implement a kind of proof reuse known as \textit{transport}~\cite{univalent2013homotopy},
but in a way that is suitable for repair.
Informally, transport takes a term $t$ and produces a term $t'$ that is the same as $t$ modulo an equivalence $A \simeq B$.
If $t$ is a function, then $t'$ behaves the same way modulo the equivalence;
if $t$ is a proof, then $t'$ proves the same theorem the same way modulo the equivalence.

When transport across $A \simeq B$ takes $t$ to $t'$,
we say that $t$ and $t'$ are \textit{equal up to transport}
across that equivalence (denoted $t \equiv_{A \simeq B} t'$).\footnote{This notation should be interpreted in a metatheory with \textit{univalence}---a property that Coq lacks---or it should be approximated in Coq.
The details of transport with univalence are in \citet{univalent2013homotopy}, and an approximation in Coq is in \citet{tabareau2017equivalences}. For equivalent \A and \B, there can be many equivalences $A \simeq B$.
Equality up to transport is across a \textit{particular} equivalence, but we erase this in the 
notation.}
In Section~\ref{sec:overview}, the original append function \lstinline{++} over \lstinline{Old.list}
and the repaired append function \lstinline{++} over \lstinline{New.list} that \toolname produces are
equal up to transport across the equivalence from Figure~\ref{fig:equivalence}, since (by \lstinline{app_ok}~\href{https://github.com/uwplse/pumpkin-pi/blob/v2.0.0/plugin/coq/Swap.v}{\circled{1}}):

\begin{lstlisting}
$\forall$ T (l1 l2 : Old.list T),(@\vspace{-0.04cm}@)
  swap T (l1 ++ l2) = (swap T l1) ++ (swap T l2).(@\vspace{-0.05cm}@)
\end{lstlisting}
The original \lstinline{rev_app_distr} is equal to the repaired proof up to transport,
since both prove the same thing the same way up to the equivalence, and up to the changes in \lstinline{++}
and \lstinline{rev}.

Transport typically works by applying the functions that make up the equivalence to convert
inputs and outputs between types.
This approach would not be suitable for repair, since it does not make it possible to remove the old type \A.
\toolname implements transport in a way that allows for removing references to \A---by proof term transformation.

%The goal of a proof repair tool like \toolname is to define a transport method that
%can remove references to the old specification, %rather than converting back and forth like standard transport methods.
%%That way, the proof repair tool can produce proofs that no longer refer in any way to the old specification,
%since the old specification may no longer exist.

%Section~\ref{sec:overview} showed a simple case of this: \toolname
%reused the proof of \lstinline{rev_app_distr} defined over \lstinline{Old.list}
%to generate a new proof of \lstinline{rev_app_distr} defined over equivalent \lstinline{New.list}.
%Furthermore, it did so in a way that removed all references to \lstinline{Old.list} in the proof
%and in its dependencies.
%That way, after calling \lstinline{Repair}, \lstinline{Old.list} could be removed.

%\subsection{A Tool for Proof Repair Across Equivalences}
%\label{sec:time}

\begin{figure}
\begin{minipage}{0.48\columnwidth}
\begin{lstlisting}[firstline=1, lastline=3]
Inductive I :=(@\vspace{-0.04cm}@)
| A : I(@\vspace{-0.04cm}@)
| B : I.(@\vspace{-0.04cm}@)
\end{lstlisting}
\end{minipage}
\hfill
\begin{minipage}{0.48\columnwidth}
\begin{lstlisting}[firstline=5, lastline=7]
Inductive J :=(@\vspace{-0.04cm}@)
| (@\codediff{makeJ}@) : (@\codediff{bool}@) $\rightarrow$ J.(@\vspace{-0.04cm}@)
$\phantom{hi}$(@\vspace{-0.04cm}@)
\end{lstlisting}
\end{minipage}
\vspace{-0.4cm}
\caption{The old type \lstinline{I} (left) is either \lstinline{A} or \lstinline{B}. The new type \lstinline{J} (right) is \lstinline{I} with \lstinline{A} and \lstinline{B} factored out to \lstinline{bool} (\codediff{orange}).}
\label{fig:equivalence2}
\end{figure}

\begin{figure}
\begin{minipage}{0.48\textwidth}
\begin{lstlisting}[firstline=1, lastline=3]
Inductive list (T : Type) : Type :=(@\vspace{-0.04cm}@)
| nil : list T(@\vspace{-0.04cm}@)
| cons : T $\rightarrow$ list T $\rightarrow$ list T.(@\vspace{-0.04cm}@)
\end{lstlisting}
\end{minipage}
\hfill
\begin{minipage}{0.58\textwidth}
\begin{lstlisting}[firstline=5, lastline=7]
Inductive vector (T : Type) : (@\codediff{nat}@) $\rightarrow$ Type :=(@\vspace{-0.04cm}@)
| nil : vector T (@\codediff{O}@)(@\vspace{-0.04cm}@)
| cons : T $\hspace{-0.06cm}\rightarrow$ $\hspace{-0.06cm}\forall\hspace{-0.06cm}$ (@\codediff{(n : nat)}@), $\hspace{-0.06cm}$vector$\hspace{-0.06cm}$ T$\hspace{-0.06cm}$ (@\codediff{n}@) $\hspace{-0.06cm}\rightarrow\hspace{-0.06cm}$ vector $\hspace{-0.06cm}$T $\hspace{-0.06cm}$(@\codediff{(S n)}@).(@\vspace{-0.04cm}@)
\end{lstlisting}
\end{minipage}
\vspace{-0.4cm}
\caption{A vector (bottom) is a list (top) indexed by its length (\codediff{orange}). Vectors effectively make it possible to enforce length invariants about lists at compile time.}
\label{fig:listtovect}
\end{figure}

\iffalse
\begin{figure*}
\begin{minipage}{0.40\textwidth}
\begin{lstlisting}[firstline=1, lastline=3]
Inductive list (T : Type) : Type :=(@\vspace{-0.04cm}@)
| nil : list T(@\vspace{-0.04cm}@)
| cons : T $\rightarrow$ list T $\rightarrow$ list T.(@\vspace{-0.04cm}@)
\end{lstlisting}
\end{minipage}
\hfill
\begin{minipage}{0.58\textwidth}
\begin{lstlisting}[firstline=5, lastline=7]
Inductive vector (T : Type) : (@\codediff{nat}@) $\rightarrow$ Type :=(@\vspace{-0.04cm}@)
| nil : vector T (@\codediff{O}@)(@\vspace{-0.04cm}@)
| cons : T $\hspace{-0.06cm}\rightarrow$ $\hspace{-0.06cm}\forall\hspace{-0.06cm}$ (@\codediff{(n : nat)}@), $\hspace{-0.06cm}$vector$\hspace{-0.06cm}$ T$\hspace{-0.06cm}$ (@\codediff{n}@) $\hspace{-0.06cm}\rightarrow\hspace{-0.06cm}$ vector $\hspace{-0.06cm}$T $\hspace{-0.06cm}$(@\codediff{(S n)}@).(@\vspace{-0.04cm}@)
\end{lstlisting}
\end{minipage}
\vspace{-0.4cm}
\caption{A vector (right) is a list (left) indexed by its length.}
\label{fig:listtovect}
\end{figure*}
\fi

\begin{figure*}
\small
\begin{grammar}
<i> $\in \mathrm{I\!N}$, <v> $\in$ Vars, <s> $\in$ \{ Prop, Set, Type<i> \}

<t> ::= <v> \hspace{0.06cm} | \hspace{0.06cm} <s> \hspace{0.06cm} | \hspace{0.06cm} $\Pi$ (<v> : <t>) . <t> \hspace{0.06cm} | \hspace{0.06cm} $\lambda$ (<v> : <t>) . <t> \hspace{0.06cm} | \hspace{0.06cm} <t> <t> \hspace{0.06cm} | \hspace{0.06cm} Ind (<v> : <t>)\{<t>,\ldots,<t>\} \hspace{0.06cm} | \hspace{0.06cm} Constr (<i>, <t>) \hspace{0.06cm} | \hspace{0.06cm} Elim(<t>, <t>)\{<t>,\ldots,<t>\}
\end{grammar}
\vspace{-0.3cm}
\caption{Syntax for CIC$_\omega$ from \citet{Timany2015FirstST} with (from left to right) variables, sorts, dependent types, functions, application, inductive types, inductive constructors, and primitive eliminators.}
\label{fig:syntax}
\end{figure*}

\section{The Transformation}
\label{sec:key2}

\begin{figure*}
\begin{minipage}{0.48\textwidth}
\begin{lstlisting}
DepConstr(0, list T) : list T := Constr((@\codediff{0}@), list T).(@\vspace{-0.04cm}@)
DepConstr(1, list T) t l : list T :=(@\vspace{-0.04cm}@)
  Constr ((@\codediff{1}@), list T) t l.(@\vspace{-0.04cm}@)
(@\vspace{-0.14cm}@)
DepElim(l, P) { p$_{\mathtt{nil}}$, p$_{\mathtt{cons}}$ } : P l :=(@\vspace{-0.04cm}@)
  Elim(l, P) { (@\codediff{p$_{\mathtt{nil}}$}@), (@\codediff{p$_{\mathtt{cons}}$}@) }.(@\vspace{-0.04cm}@)
\end{lstlisting}
\end{minipage}
\hfill
\begin{minipage}{0.48\textwidth}
\begin{lstlisting}
DepConstr(0, list T) : list T := Constr((@\codediff{1}@), list T).(@\vspace{-0.04cm}@)
DepConstr(1, list T) t l : list T :=(@\vspace{-0.04cm}@)
  Constr((@\codediff{0}@), list T) t l.(@\vspace{-0.04cm}@)
(@\vspace{-0.14cm}@)
DepElim(l, P) { p$_{\mathtt{nil}}$, p$_{\mathtt{cons}}$ } : P l :=(@\vspace{-0.04cm}@)
  Elim(l, P) { (@\codediff{p$_{\mathtt{cons}}$}@), (@\codediff{p$_{\mathtt{nil}}$}@) }.(@\vspace{-0.04cm}@)
\end{lstlisting}
\end{minipage}
\vspace{-0.3cm}
\caption{The dependent constructors and eliminators for old (left) and new (right) \lstinline{list}, with the difference in \codediff{orange}.}
\vspace{-0.1cm}
\label{fig:listconfig}
\end{figure*}

At the heart of \toolname is a configurable proof term transformation for transporting
proofs across equivalences~\href{https://github.com/uwplse/pumpkin-pi/blob/v2.0.0/plugin/src/automation/lift/lift.ml}{\circled{4}}. % lift.ml
It is a generalization of the transformation from an earlier version of \toolname called
\textsc{Devoid}~\cite{Ringer2019}, which solved this problem a particular class of equivalences.
%\toolname moves the reasoning specific to that class of equivalences into the configuration. 

%\begin{quote}
%\textbf{Insight 2}:
%A configurable proof term transformation can be used to build such a proof repair tool,
%and the result can handle many different kinds of changes.
%\end{quote}

The transformation takes as input a deconstructed equivalence that we call a \textit{configuration}.
This section introduces the configuration (Section~\ref{sec:configurable}),
defines the transformation that builds on that (Section~\ref{sec:generic}),
then specifies correctness criteria for the configuration (Section~\ref{sec:art}).
Section~\ref{sec:implementation} describes the additional work needed to implement this transformation.

\mysubsubsec{Conventions}
All terms that we introduce in this section are in the Calculus of Inductive Constructions (CIC$_{\omega}$), the type theory
that Coq's proof term language Gallina implements.
CIC$_{\omega}$ is based on the Calculus of Constructions (CoC), a variant of the lambda calculus with polymorphism (types that depend on types) and dependent types (types that depend on terms)~\cite{coquand:inria-00076024}. CIC$_{\omega}$ extends CoC with
inductive types~\cite{inductive}.
Inductive types are defined solely by their constructors (like \lstinline{nil} and \lstinline{cons} for \lstinline{list}) and eliminators (like the induction principle for \lstinline{list}); this section assumes that these eliminators are primitive.

The syntax for CIC$_{\omega}$ with primitive eliminators is in Figure~\ref{fig:syntax}.
The typing rules are standard.
We assume inductive types $\Sigma$ with constructor $\exists$ and projections $\pi_l$ and $\pi_r$,
and an equality type \lstinline{=} with constructor \lstinline{eq_refl}.
We use $\vec{t}$ and $\{t_1, \ldots, t_n\}$ to denote lists of terms.

\subsection{The Configuration}
\label{sec:configurable}

The configuration is the key to building a proof term transformation that implements transport in a way that is suitable for repair.
Each configuration corresponds to an equivalence \A $\simeq$ \B.
It deconstructs the equivalence into things that talk about \A, and things that talk about \B.
It does so in a way that hides details
specific to the equivalence, like the order or number of arguments to an induction principle or type.

At a high level, the configuration helps the transformation achieve two goals: preserve equality up to transport across the equivalence 
between \A and \B, and produce well-typed terms.
This configuration is a pair of pairs:

\begin{lstlisting}
((DepConstr, DepElim), (Eta, Iota))(@\vspace{-0.05cm}@)
\end{lstlisting}
each of which corresponds to one of the two goals:
\lstinline{DepConstr} and \lstinline{DepElim} define how to transform constructors and eliminators, thereby preserving the equivalence, and 
\lstinline{Eta} and \lstinline{Iota} define how to transform $\eta$-expansion and $\iota$-reduction of constructors and eliminators, thereby producing well-typed terms.
Each of these is defined in CIC$_{\omega}$ for each equivalence.
%\textbf{Configure} passes this configuration to \textbf{Transform}.

%Section~\ref{sec:art} describes how the four parts of this configuration must relate to one another in order for the proof
%term transformation to work correctly, and proves that every equivalence induces a configuration.

\mysubsubsec{Preserving the Equivalence}
To preserve the equivalence, the configuration ports terms over \A to terms over \B by viewing each
term of type \B as if it were an \A.
This way, the rest of the transformation can replace values of \A with values of \B, and
inductive proofs about \A with inductive proofs about \B, %, then recursively transform
%subterms 
all without changing the order or number of arguments.

The configuration parts responsible for this are \lstinline{DepConstr}
and \lstinline{DepElim} (\textit{dependent constructors} and \textit{eliminators}).
These describe how to construct and eliminate \A and \B, wrapping the types with a common inductive structure.
The transformation requires the same number of dependent constructors and cases in dependent eliminators for \A and \B,
even if \A and \B are types with different numbers of constructors
(\A and \B need not even be inductive; see Sections~\ref{sec:art} and~\ref{sec:search}).

For the \lstinline{list} change from Section~\ref{sec:overview},
the configuration that \toolname discovers uses the dependent constructors
and eliminators in Figure~\ref{fig:listconfig}. The dependent constructors for \lstinline{Old.list}
are the normal constructors with the order unchanged,
while the dependent constructors for \lstinline{New.list} swap constructors
back to the original order.
Similarly, the dependent eliminator for \lstinline{Old.list} is the normal eliminator for \lstinline{Old.list},
while the dependent eliminator for \lstinline{New.list} swaps cases.

As the name hints, these constructors and eliminators can be dependent.
Consider the type of vectors of some length:

\begin{lstlisting}
packed_vect T := $\Sigma$(n : nat).vector T n.(@\vspace{-0.05cm}@)
\end{lstlisting}
\toolname can port proofs across the equivalence between this type and \lstinline{list T}~\href{https://github.com/uwplse/pumpkin-pi/blob/v2.0.0/plugin/coq/examples/Example.v}{\circled{3}}. % Example.v
The dependent constructors \toolname discovers pack the index into an existential, like:

\begin{lstlisting}
DepConstr(0, packed_vect) : packed_vect T :=(@\vspace{-0.04cm}@)
  $\exists$ (Constr(0, nat)) (Constr(0, vector T)).(@\vspace{-0.05cm}@)
\end{lstlisting}
and the eliminator it discovers eliminates the projections:

\begin{lstlisting}
DepElim(s, P) { f$_0$ f$_1$ } : P ($\exists$ ($\pi_l$ s) ($\pi_r$ s)) :=(@\vspace{-0.04cm}@)
  Elim($\pi_r$ s, $\lambda$(n : nat)(v : vector T n).P ($\exists$ n v)) {(@\vspace{-0.04cm}@)
    f$_0$,(@\vspace{-0.04cm}@)
    ($\lambda$(t : T)(n : nat)(v : vector T n).f$_1$ t ($\exists$ n v))(@\vspace{-0.04cm}@)
  }.(@\vspace{-0.05cm}@) 
\end{lstlisting}

In both these examples, the interesting work moves into the configuration:
the configuration for the first swaps constructors and cases,
and the configuration for the second maps constructors and cases over \lstinline{list} to constructors and cases over \lstinline{packed_vect}. %packs constructors and eliminates projections.
That way, the transformation need not add, drop, or reorder arguments.
%In essence, all of the difficult work moves into the configuration.
Furthermore, both examples use automatic configuration, so \toolname's \textbf{Configure} component 
discovers \lstinline{DepConstr} and \lstinline{DepElim} from just the types \A and \B, taking care of even the difficult work.

\mysubsubsec{Producing Well-Typed Terms}
The other configuration parts \lstinline{Eta} and \lstinline{Iota} deal with producing well-typed terms,
in particular by transporting equalities.
CIC$_{\omega}$ distinguishes between two important kinds of equality: those that hold by reduction (\textit{definitional} equality), and those that hold by proof (\textit{propositional} equality).
That is, two terms \lstinline{t} and \lstinline{t'} of type \lstinline{T} are definitionally equal if they reduce to the same normal form,
and propositionally equal if there is a proof that \lstinline{t = t'} using the inductive
equality type \lstinline{=} at type \lstinline{T}. Definitionally equal terms are necessarily propositionally equal, but 
the converse is not in general true.

When a datatype changes, sometimes, definitional equalities defined over the old version of that type must become propositional.
A naive proof term transformation may fail to generate well-typed terms if it does not account for this.
Otherwise, if the transformation transforms a term \lstinline{t : T} to some \lstinline{t' : T'}, it does not necessarily
transform \lstinline{T} to \lstinline{T'}~\cite{tabareau2019marriage}.

\lstinline{Eta} and \lstinline{Iota} describe how to transport equalities.
More formally, they define $\eta$-expansion and $\iota$-reduction of \A and \B,
which may be propositional rather than definitional,
and so must be explicit in the transformation.
$\eta$-expansion describes how to expand a term to apply a constructor to an eliminator in a way that preserves propositional equality,
and is important for defining dependent eliminators~\cite{nlab:eta-conversion}.
$\iota$-reduction ($\beta$-reduction for inductive types) describes how to reduce an elimination of a constructor~\cite{nlab:beta-reduction}.

The configuration for the change from \lstinline{list} to \lstinline{packed_vect} has propositional \lstinline{Eta}.
It uses $\eta$-expansion for $\Sigma$:

\begin{lstlisting}
Eta(packed_vect) := $\lambda$(s:packed_vect).$\exists$ ($\pi_l$ s) ($\pi_r$ s).(@\vspace{-0.05cm}@)
\end{lstlisting}
which is propositional and not definitional in Coq.
Thanks to this, we can forego the assumption that our language has primitive projections (definitional $\eta$ for $\Sigma$).

\begin{figure}
\begin{minipage}{0.44\columnwidth}
\begin{lstlisting}[firstline=1, lastline=8]
(@\vspace{-0.04cm}@)
(@\vspace{-0.04cm}@)
(@\vspace{-0.04cm}@)
(@\vspace{-0.04cm}@)
(@\vspace{-0.14cm}@)
Inductive nat :=(@\vspace{-0.04cm}@)
| O : nat(@\vspace{-0.04cm}@)
| S : nat $\rightarrow$ nat.(@\vspace{-0.04cm}@)
\end{lstlisting}
\end{minipage}
\hfill
\begin{minipage}{0.54\columnwidth}
\begin{lstlisting}[firstline=10, lastline=17]
Inductive positive :=(@\vspace{-0.04cm}@)
| xI : positive  $\rightarrow$ positive(@\vspace{-0.04cm}@)
| xO : positive  $\rightarrow$ positive (@\vspace{-0.04cm}@)
| xH : positive.(@\vspace{-0.04cm}@)
(@\vspace{-0.14cm}@)
Inductive N :=(@\vspace{-0.04cm}@)
| N0 : N (@\vspace{-0.04cm}@)
| Npos : (@\codediff{positive}@) $\rightarrow$ N.(@\vspace{-0.04cm}@)
\end{lstlisting}
\end{minipage}
\vspace{-0.2cm}
\caption{A unary natural number \lstinline{nat} (left) is either zero (\lstinline{0}) or the successor of some other natural number (\lstinline{S}).
A binary natural number \lstinline{N} (right) is either zero (\lstinline{N0}) or a positive binary number (\lstinline{Npos}), where a positive binary number is either 1 (\lstinline{xH}), or the result of shifting left and adding 1 (\lstinline{xI}) or
0 (\lstinline{xO}). Unary and binary natural numbers are equivalent, but have different inductive structures.
Consequentially, definitional equalities over \lstinline{nat} may become propositional over \lstinline{N}.}
\vspace{-0.2cm}
\label{fig:nattobin}
\end{figure}

Each \lstinline{Iota}---one per constructor---describes and proves the $\iota$-reduction behavior
of \lstinline{DepElim} on the corresponding case.
This is needed, for example, to port proofs about unary numbers \lstinline{nat} to
proofs about binary numbers \lstinline{N} (Figure~\ref{fig:nattobin}).
While we can define \lstinline{DepConstr} and \lstinline{DepElim} to induce an equivalence
between them~\href{https://github.com/uwplse/pumpkin-pi/blob/v2.0.0/plugin/coq/nonorn.v}{\circled{5}}, % nonorn.v
we run into trouble reasoning about applications of \lstinline{DepElim},
since proofs about \lstinline{nat} that hold by reflexivity do not necessarily hold by reflexivity over \lstinline{N}. 
For example, in Coq, while \lstinline{S (n + m)  = S n + m} holds by reflexivity over \lstinline{nat},
when we define \lstinline{+} with \lstinline{DepElim} over \lstinline{N},
the corresponding theorem over \lstinline{N} does not hold by reflexivity.

To transform proofs about \lstinline{nat} to proofs about \lstinline{N}, we must transform \textit{definitional} $\iota$-reduction over \lstinline{nat} to \textit{propositional} $\iota$-reduction over \lstinline{N}.
For our choice of \lstinline{DepConstr} and \lstinline{DepElim},
$\iota$-reduction is definitional over \lstinline{nat}, since a proof of:

\begin{lstlisting}
$\forall$ P p$_\texttt{0}$ p$_\texttt{S}$ n,(@\vspace{-0.04cm}@)
  DepElim((@\codediff{DepConstr(1, nat) n}@), P) { p$_\texttt{0}$, p$_\texttt{S}$ } =(@\vspace{-0.04cm}@)
  (@\codediff{p$_\texttt{S}$}@) n (DepElim(n, P) { p$_\texttt{0}$, p$_\texttt{S}$ }).(@\vspace{-0.05cm}@)
\end{lstlisting}
holds by reflexivity.
\lstinline{Iota} for \lstinline{nat} in the \lstinline{S} case is a rewrite by that proof by reflexivity~\href{https://github.com/uwplse/pumpkin-pi/blob/v2.0.0/plugin/coq/nonorn.v}{\circled{5}},
with type:

\begin{lstlisting}
$\forall$ P p$_\texttt{0}$ p$_\texttt{S}$ n (Q: P (DepConstr(1, nat) n) $\rightarrow$ s),(@\vspace{-0.04cm}@)
  Iota(1, nat, Q) :(@\vspace{-0.04cm}@)
    Q ((@\codediff{p$_\texttt{S}$}@) n (DepElim(n, P) { p$_\texttt{0}$, p$_\texttt{S}$ })) $\rightarrow$(@\vspace{-0.04cm}@)
    Q (DepElim((@\codediff{DepConstr(1, nat) n}@), P) { p$_\texttt{0}$, p$_\texttt{S}$ }).(@\vspace{-0.05cm}@)
\end{lstlisting}
In contrast, $\iota$ for \lstinline{N} is propositional, since the 
theorem: %over \lstinline{N}:

\begin{lstlisting}
$\forall$ P p$_\texttt{0}$ p$_\texttt{S}$ n,(@\vspace{-0.04cm}@)
  DepElim((@\codediff{DepConstr(1, N) n}@), P) { p$_\texttt{0}$, p$_\texttt{S}$ } =(@\vspace{-0.04cm}@)
  (@\codediff{p$_\texttt{S}$}@) n (DepElim(n, P) { p$_\texttt{0}$, p$_\texttt{S}$ }).(@\vspace{-0.05cm}@)
\end{lstlisting}
no longer holds by reflexivity.
\lstinline{Iota} for \lstinline{N} is a rewrite by the propositional equality that proves this theorem~\href{https://github.com/uwplse/pumpkin-pi/blob/v2.0.0/plugin/coq/nonorn.v}{\circled{5}},
with type:

\begin{lstlisting}
$\forall$ P p$_\texttt{0}$ p$_\texttt{S}$ n (Q: P (DepConstr(1, N) n) $\rightarrow$ s),(@\vspace{-0.04cm}@)
  Iota(1, N, Q) :(@\vspace{-0.04cm}@)
    Q ((@\codediff{p$_\texttt{S}$}@) n (DepElim(n, P) { p$_\texttt{0}$, p$_\texttt{S}$ })) $\rightarrow$(@\vspace{-0.04cm}@)
    Q (DepElim((@\codediff{DepConstr(1, N) n}@), P) { p$_\texttt{0}$, p$_\texttt{S}$ }).(@\vspace{-0.05cm}@)
\end{lstlisting}
By replacing \lstinline{Iota} over \lstinline{nat} with \lstinline{Iota} over \lstinline{N},
the transformation replaces rewrites by reflexivity over \lstinline{nat} to rewrites by propositional equalities over \lstinline{N}.
That way, \lstinline{DepElim} behaves the same over \lstinline{nat} and \lstinline{N}.

Taken together over both \A and \B, \lstinline{Iota} describes how the inductive structures of \A and \B differ.
The transformation requires that \lstinline{DepElim} over \A and over \B have the same structure
as each other, so if \A and \B \textit{themselves} have the same 
inductive structure (if they are \textit{ornaments}~\cite{mcbride}),
then if $\iota$ is definitional for \A, it will be possible to choose
\lstinline{DepElim} with definitional $\iota$ for \B.
Otherwise, if \A and \B (like \lstinline{nat} and \lstinline{N}) have different inductive structures,
then definitional $\iota$ over one would become propositional $\iota$ over the other.
%For the case of \lstinline{nat} and \lstinline{N},
%the need for propositional $\iota$ was noted as far back as \citet{magaud2000changing}.
%\lstinline{Iota} in the configuration encodes this more generally.

\subsection{The Proof Term Transformation}
\label{sec:generic}

\begin{figure*}
\begin{mathpar}
\mprset{flushleft}
\small
\hfill\fbox{$\Gamma$ $\vdash$ $t$ $\Uparrow$ $t'$}\vspace{-0.3cm}\\

\inferrule[Dep-Elim]
  { \Gamma \vdash a \Uparrow b \\ \Gamma \vdash p_{a} \Uparrow p_b \\ \Gamma \vdash \vec{f_{a}}\phantom{l} \Uparrow \vec{f_{b}} }
  { \Gamma \vdash \mathrm{DepElim}(a,\ p_{a}) \vec{f_{a}} \Uparrow \mathrm{DepElim}(b,\ p_b) \vec{f_{b}} }

\inferrule[Dep-Constr]
{ \Gamma \vdash \vec{t}_{a} \Uparrow \vec{t}_{b} } %\\ TODO must we explicitly lift A to B if we want to handle parameters/indices?
{ \Gamma \vdash \mathrm{DepConstr}(j,\ A)\ \vec{t}_{a} \Uparrow \mathrm{DepConstr}(j,\ B)\ \vec{t}_{b}  }

\inferrule[Eta]
  { \\ }
  { \Gamma \vdash \mathrm{Eta}(A) \Uparrow \mathrm{Eta}(B) }

\inferrule[Iota]
  { \Gamma \vdash q_A \Uparrow q_B \\ \Gamma \vdash \vec{t_A} \Uparrow \vec{t_B} }
  { \Gamma \vdash \mathrm{Iota}(j,\ A,\ q_A)\ \vec{t_A} \Uparrow \mathrm{Iota}(j,\ B,\ q_B)\ \vec{t_B} }

\inferrule[Equivalence]
  { \\ }
  { \Gamma \vdash A\ \Uparrow B }

\inferrule[Constr]
{ \Gamma \vdash T \Uparrow T' \\ \Gamma \vdash \vec{t} \Uparrow \vec{t'} }
{ \Gamma \vdash \mathrm{Constr}(j,\ T)\ \vec{t} \Uparrow \mathrm{Constr}(j,\ T')\ \vec{t'} }

\inferrule[Ind]
  { \Gamma \vdash T \Uparrow T' \\ \Gamma \vdash \vec{C} \Uparrow \vec{C'}  }
  { \Gamma \vdash \mathrm{Ind} (\mathit{Ty} : T) \vec{C} \Uparrow \mathrm{Ind} (\mathit{Ty} : T') \vec{C'} }

%% Application
\inferrule[App]
 { \Gamma \vdash f \Uparrow f' \\ \Gamma \vdash t \Uparrow t'}
 { \Gamma \vdash f t \Uparrow f' t' }

\inferrule[Elim] % TODO wait why do we have c here when it clearly refers to the term we eliminate over? um
  { \Gamma \vdash c \Uparrow c' \\ \Gamma \vdash Q \Uparrow Q' \\ \Gamma \vdash \vec{f} \Uparrow \vec{f'}}
  { \Gamma \vdash \mathrm{Elim}(c, Q) \vec{f} \Uparrow \mathrm{Elim}(c', Q') \vec{f'}  }

% Lamda
\inferrule[Lam]
  { \Gamma \vdash t \Uparrow t' \\ \Gamma \vdash T \Uparrow T' \\ \Gamma,\ t : T \vdash b \Uparrow b' }
  {\Gamma \vdash \lambda (t : T).b \Uparrow \lambda (t' : T').b'}

% Product
\inferrule[Prod]
  { \Gamma \vdash t \Uparrow t' \\ \Gamma \vdash T \Uparrow T' \\ \Gamma,\ t : T \vdash b \Uparrow b' }
  {\Gamma \vdash \Pi (t : T).b \Uparrow \Pi (t' : T').b'}

\inferrule[Var]
  { v \in \mathrm{Vars} }
  {\Gamma \vdash v \Uparrow v}

%\inferrule[Sort]
%  { \\ }
%  {\Gamma \vdash s \Uparrow s}
\end{mathpar}
\vspace{-0.3cm}
\caption{Transformation for transporting terms across $A \simeq B$ with configuration \lstinline{((DepConstr, DepElim), (Eta, Iota))}.}
\label{fig:final}
\end{figure*}

\begin{figure*}
\begin{minipage}{0.49\textwidth}
\begin{lstlisting}
(* 1: original term *)(@\vspace{-0.04cm}@)
$\lambda$ (T : Type) (l m : Old.list T) .(@\vspace{-0.04cm}@)
 Elim(l, $\lambda$(l: Old.list T).Old.list T $\rightarrow$ Old.list T)) {(@\vspace{-0.04cm}@)
   ($\lambda$ m . m),(@\vspace{-0.04cm}@)
   ($\lambda$ t _ IHl m . Constr(1, Old.list T) t (IHl m))(@\vspace{-0.04cm}@)
 } m.(@\vspace{-0.04cm}@)
(@\vspace{-0.10cm}@)
(* 2: after unifying (@\texttt{with}@) configuration *)(@\vspace{-0.04cm}@)
$\lambda$ (T : Type) (l m : (@\codediff{A}@)) .(@\vspace{-0.04cm}@)
 (@\codediff{DepElim}@)(l, $\lambda$(l: (@\codediff{A}@)).(@\codediff{A}@) $\rightarrow$ (@\codediff{A}@))) {(@\vspace{-0.04cm}@)
   ($\lambda$ m . m)(@\vspace{-0.04cm}@)
   ($\lambda$ t _ IHl m . (@\codediff{DepConstr}@)(1, (@\codediff{A}@)) t (IHl m))(@\vspace{-0.04cm}@)
 } m.(@\vspace{-0.04cm}@)
\end{lstlisting}
\end{minipage}
\hfill
\begin{minipage}{0.49\textwidth}
\begin{lstlisting}
(* 4: reduced to final term *)(@\vspace{-0.04cm}@)
$\lambda$ (T : Type) (l m : New.list T) .(@\vspace{-0.04cm}@)
 Elim(l, $\lambda$(l: New.list T).New.list T $\rightarrow$ New.list T)) {(@\vspace{-0.04cm}@)
   ($\lambda$ t _ IHl m . Constr(0, New.list T) t (IHl m)),(@\vspace{-0.04cm}@)
   ($\lambda$ m . m)(@\vspace{-0.04cm}@)
 } m.(@\vspace{-0.04cm}@)
(@\vspace{-0.10cm}@)
(* 3: after transforming *)(@\vspace{-0.04cm}@)
$\lambda$ (T : Type) (l m : (@\codediff{B}@)) .(@\vspace{-0.04cm}@)
 (@\codediff{DepElim}@)(l, $\lambda$(l: (@\codediff{B}@)).(@\codediff{B}@) $\rightarrow$ (@\codediff{B}@))) {(@\vspace{-0.04cm}@)
   ($\lambda$ m . m)(@\vspace{-0.04cm}@)
   ($\lambda$ t _ IHl m . (@\codediff{DepConstr}@)(1, (@\codediff{B}@)) t (IHl m))(@\vspace{-0.04cm}@)
 } m.(@\vspace{-0.04cm}@)
\end{lstlisting}
\end{minipage}
\vspace{-0.3cm}
\caption{Swapping cases of the append function, counterclockwise, the input term: 1) unmodified, 2) unified with the configuration, 3) ported to the updated type, and 4) reduced to the output.}
\label{fig:appswap1}
\end{figure*}

Figure~\ref{fig:final} shows the proof term transformation $\Gamma \vdash t \Uparrow t'$ that forms the core of \toolname.
%Like the transformation from \textsc{Devoid},
The transformation is parameterized over equivalent types \A and \B (\textsc{Equivalence})
as well as the configuration. %terms, which appear in the transformation explicitly.
It assumes $\eta$-expanded functions.
It implicitly constructs an updated context $\Gamma'$ in which to interpret $t'$, but this is not needed for computation.

The proof term transformation is (perhaps deceptively) simple by design:
it moves the bulk of the work into the configuration,
and represents the configuration explicitly.
Of course, typical proof terms in Coq do not apply these configuration
terms explicitly.
\toolname does some additional work using \textit{unification heuristics} to get real proof terms into this format before running the transformation.
It then runs the proof term transformation, which transports proofs across the equivalence that corresponds to the configuration.

\mysubsubsec{Unification Heuristics}
The transformation does not fully describe the search procedure for transforming terms that \toolname implements.
Before running the transformation, \toolname \textit{unifies} subterms with particular \A (fixing parameters and indices),
and with applications of configuration terms over \A. 
The transformation then transforms configuration terms over \A
to configuration terms over \B.
Reducing the result produces the output term defined over \B.

Figure~\ref{fig:appswap1} shows this with the list append function \lstinline{++} from Section~\ref{sec:overview}.
To update \lstinline{++} (top left), \toolname unifies \lstinline{Old.list T} with \A, and \lstinline{Constr} and \lstinline{Elim}
with \lstinline{DepConstr} and \lstinline{DepElim} (bottom left).
After unification, the transformation recursively substitutes \B
for \A, which moves \lstinline{DepConstr} and \lstinline{DepElim}
to construct and eliminate over the updated type (bottom right).
This reduces to a term with swapped constructors and cases over \lstinline{New.list T} (top right).

In this case, unification is straightforward. % since \lstinline{DepConstr} and \lstinline{DepElim} correspond to
%\lstinline{Constr} and \lstinline{Elim} directly.
This can be more challenging when configuration terms are dependent.
This is especially pronounced with definitional \lstinline{Eta} and \lstinline{Iota},
which typically are implicit (reduced) in real code.
%This problem is exactly why \citet{tabareau2019marriage} speculated that converting definitional to propositional equalities
%like we do with \lstinline{Iota} may, in general, be intractable.
To handle this, \toolname implements custom \textit{unification heuristics} for each search procedure
that unify subterms with applications of configuration terms, and that instantiate parameters and dependent indices in those subterms~\href{https://github.com/uwplse/pumpkin-pi/blob/v2.0.0/plugin/src/automation/lift/liftconfig.ml}{\circled{6}}. % liftconfig.ml
The transformation in turn assumes that all existing parameters and indices are determined and instantiated
by the time it runs.

\toolname falls back to Coq's unification for manual configuration and when these custom heuristics fail.
When even Coq's unification is not enough, \toolname relies on proof engineers to provide hints
in the form of annotations~\href{https://github.com/uwplse/pumpkin-pi/blob/v2.0.0/plugin/coq/nonorn.v}{\circled{5}}.

\begin{figure*}
\begin{minipage}{0.43\textwidth}
\begin{lstlisting}
section: $\forall$ (a : A), g (f a) = a.(@\vspace{-0.04cm}@)
retraction: $\forall$ (b : B), f (g b) = b.(@\vspace{-0.04cm}@)
(@\vspace{-0.14cm}@)
constr_ok: $\forall$ $j$ $\vec{x_A}$ $\vec{x_B}$, $\vec{x_A}$ $\equiv_{A \simeq B}$ $\vec{x_B}$ $\rightarrow$(@\vspace{-0.04cm}@)
  DepConstr($j$, A) $\vec{x_A}$ $\equiv_{A \simeq B}$ DepConstr(j, B) $\vec{x_B}$.(@\vspace{-0.04cm}@)
(@\vspace{-0.14cm}@)
elim_ok: $\forall$ a b P$_A$ P$_B$ $\vec{f_A}$ $\vec{f_B}$,(@\vspace{-0.04cm}@)
  a $\equiv_{A \simeq B}$ b $\rightarrow$(@\vspace{-0.04cm}@)
  P$_A$ $\equiv_{(A \rightarrow s) \simeq (B \rightarrow s)}$ P$_B$ $\rightarrow$(@\vspace{-0.04cm}@)
  $\forall$ $j$, $\vec{f_A}$[j] $\equiv_{\xi (A, P_A, j) \simeq \xi (B, P_B, j)}$ $\vec{f_B}$[j]$\rightarrow$(@\vspace{-0.04cm}@)
  DepElim(a, P$_A$) $\vec{f_A}$ $\equiv_{(P a) \simeq (P b)}$ DepElim(b, P$_B$) $\vec{f_A}$.(@\vspace{-0.04cm}@)
\end{lstlisting}
\end{minipage}
\hfill
\begin{minipage}{0.56\textwidth}
\begin{lstlisting}
elim_eta(A): $\forall$ a P $\vec{f}$, DepElim(a, P) $\vec{f}$ : P (Eta(A) a).(@\vspace{-0.04cm}@)
eta_ok(A): $\forall$ (a : A), Eta(A) a = a.(@\vspace{-0.04cm}@)
(@\vspace{-0.14cm}@)
(@\phantom{constr_ok: $\forall$ $j$ $\vec{x_A}$ $\vec{x_B}$,}@)(@\vspace{-0.04cm}@)
(@\phantom{  DepConstr($j$, A) $\vec{x_A}$ $\equiv_{A \simeq B}$ DepConstr(j, B) $\vec{x_B}$.}@)(@\vspace{-0.04cm}@)
(@\vspace{-0.14cm}@)
iota_ok(A): $\forall$ $j$ P $\vec{f}$ $\vec{x}$ (Q: P(Eta(A) (DepConstr($j$, A) $\vec{x}$)) $\rightarrow$ s),(@\vspace{-0.04cm}@)
  Iota(A, j, Q) : (@\vspace{-0.04cm}@)
    Q (DepElim(DepConstr(j, A) $\vec{x}$, P) $\vec{f}$) $\rightarrow$ (@\vspace{-0.04cm}@)
    Q (rew $\leftarrow$ eta_ok(A) (DepConstr(j, A) $\vec{x}$) in(@\vspace{-0.04cm}@)
      ($\vec{f}$[j]$\ldots$(DepElim(IH$_0$, P) $\vec{f}$)$\ldots$(DepElim(IH$_n$, P) $\vec{f}$)$\ldots$)).(@\vspace{-0.04cm}@)
\end{lstlisting}
\end{minipage}
\vspace{-0.2cm}
\caption{Correctness criteria for a configuration to ensure that the transformation
preserves equivalence (left) coherently with equality (right, shown for \A; \B is similar). \lstinline{f} and \lstinline{g} are defined in text. $s$, $\vec{f}$, $\vec{x}$, and $\vec{\mathtt{IH}}$ represent
sorts, eliminator cases, constructor arguments, and inductive hypotheses. $\xi$ $(A,$ $P,$ $j)$ is the type 
of \lstinline{DepElim(A, P)} at \lstinline{DepConstr(j, A)} (similarly for \B).} %, respectively.}
\label{fig:spec}
\end{figure*}
 % TODO sigs for Iota here are not quite correct---Q is not bound. Also need to tweak to deal w/ eta, and to use eta_OK, and to relate both eta
% TODO (!!) Define f and g with some schema like:
% ж(A, B) := λ(a : A).DepElim(a, $\lambda$(a : A).B){ $\lambda$ ... DepConstr(0, B) ..., ... }
% ж(A, B) := λ(b : B).DepElim(b, $\lambda$(b : B).A){ $\lambda$ ... DepConstr(0, A) ..., ... }

% $\mathrm{E}_{A_i}\ (p_A : \mathrm{P}_A)$ := $\xi(A,\ p_A,\ \mathrm{Constr}(i,\ A),\ C_{A_i})$

\mysubsubsec{Specifying a Correct Transformation}
The implementation of this transformation in \toolname produces a term that Coq type checks, and so does not
add to the trusted computing base.
As \toolname is an engineering tool, there is no need to formally prove the transformation correct, though doing so would be satisfying.
The goal of such a proof would be to show that % the transformation preserves equality up to transport along the equivalence $A \simeq B$,
%while no longer referring to the old specification.
%That is, we need that 
if $\Gamma \vdash t \Uparrow t'$,
then $t$ and $t'$ are equal up to transport, and $t'$ refers to \B in place of \A.
%This is the same as the correctness criterion for the program transformation from \textsc{Devoid} that this is based on,
%with the transformation generalized to handle other equivalences beyond the class that \textsc{Devoid} supports.
The key steps in this transformation that make this possible are porting terms along the configuration % corresponding
%to a particular equivalence 
(\textsc{Dep-Constr}, \textsc{Dep-Elim}, \textsc{Eta}, and \textsc{Iota}).
%The rest is straightforward.
For metatheoretical reasons, without additional axioms, a proof of this theorem in Coq can only be approximated~\cite{tabareau2017equivalences}.
It would be possible to generate per-transformation proofs of correctness, but this does not serve an engineering need.

\subsection{Specifying Correct Configurations}
\label{sec:art}

%Both when designing a search procedure for an automatic configuration and when
%configuring \toolname manually, choosing a configuration is important,
%and it is not always straightforward.
%This section specifies what it means for a configuration to be correct. % and gives some intuition as to why.
%Section~\ref{sec:search} shows some useful example configurations.
%The configuration instantiates the proof term transformation to a particular equivalence between \A and \B.

Choosing a configuration necessarily depends in some way on the proof engineer's intentions:
there can be infinitely many equivalences that correspond to a 
change, only some of which are useful (for example~\href{https://github.com/uwplse/pumpkin-pi/blob/v2.0.0/plugin/coq/playground/refine_unit.v}{\circled{7}}, any \A is equivalent to \lstinline{unit} refined by \A). % refine_unit.v
And there can be many configurations that correspond
to an equivalence, some of which will produce terms that are more useful or efficient than others
(consider \lstinline{DepElim} converting through several intermediate types).

While we cannot control for intentions, we \textit{can} specify what it means for a chosen configuration to be correct:
Fix a configuration. Let \lstinline{f} be the function that uses \lstinline{DepElim} to eliminate \A and \lstinline{DepConstr} to construct \B,
and let \lstinline{g} be similar. %Assume a univalent metatheory in which equality up to transport is defined.
Figure~\ref{fig:spec} specifies the correctness criteria for the configuration.
These criteria relate \lstinline{DepConstr}, \lstinline{DepElim}, \lstinline{Eta}, and \lstinline{Iota}
in a way that preserves equivalence coherently with equality.

\mysubsubsec{Equivalence}
To preserve the equivalence (Figure~\ref{fig:spec}, left), \lstinline{DepConstr} and \lstinline{DepElim} must form an equivalence
%between \A and \B.
(\lstinline{section} and \lstinline{retraction} must hold for \lstinline{f} and \lstinline{g}).
%uses  \lstinline{DepElim} to eliminate \B and \lstinline{DepConstr} to construct \A.
\lstinline{DepConstr} over \A and \B must be equal up to transport across that equivalence (\lstinline{constr_ok}), 
and similarly for \lstinline{DepElim} (\lstinline{elim_ok}).
%An example proves this on the change from \lstinline{list T} to \lstinline{packed_vect T} in the 
%univalent parametricity framework.\footnote{\url{https://github.com/CoqHott/univalent_parametricity/commit/7dc14e69942e6b3302fadaf5356f9a7e724b0f3c}}
Intuitively, \lstinline{constr_ok} and \lstinline{elim_ok} guarantee that the transformation
correctly transports dependent constructors and dependent eliminators,
as doing so will preserve equality up to transport for those subterms.
This makes it possible for the transformation
to avoid applying \lstinline{f} and \lstinline{g}, instead porting terms from \A directly to \B.

%Furthermore, since CIC$_{\omega}$ is constructive, the \textit{only} way to construct an \A (respectively \B) is to use its constructors,
%and the \textit{only} way to eliminate an \A (respectively \B) is to apply its eliminator.
%Finally, since these form an equivalence, all ways of constructing or eliminating \A and \B are covered by these dependent constructors and %eliminators.
%So, as long as we are able to unify subterms with applications of \lstinline{DepConstr} and \lstinline{DepElim},
%\textsc{Dep-Constr} and \textsc{Dep-Elim} should preserve correctness of the transformation and cover all values and eliminations of \A and \B.

\begin{figure*}
\small
\begin{grammar}
<v> $\in$ Vars, <t> $\in$ CIC$_{\omega}$

<p> ::= intro <v> \hspace{0.05cm} | \hspace{0.05cm} rewrite <t> <t> \hspace{0.05cm} | \hspace{0.05cm} symmetry \hspace{0.05cm} | \hspace{0.05cm} apply <t> \hspace{0.05cm} | \hspace{0.05cm} induction <t> <t> \{ <p>, \ldots, <p> \} \hspace{0.05cm} | \hspace{0.05cm} split \{ <p>, <p> \} \hspace{0.05cm} | \hspace{0.05cm} left \hspace{0.05cm} | \hspace{0.05cm} right \hspace{0.05cm} | \hspace{0.05cm} <p> . <p>
\end{grammar}
\vspace{-0.4cm}
\caption{Qtac syntax.}
\vspace{-0.4cm}
\label{fig:ltacsyntax1}
\end{figure*}

\begin{figure*}
\begin{mathpar}
\mprset{flushleft}
\small
\hfill\fbox{$\Gamma$ $\vdash$ $t$ $\Rightarrow$ $p$}\vspace{-0.5cm}\\

\inferrule[Intro]
  { \Gamma,\ n : T \vdash b \Rightarrow p }
  { \Gamma \vdash \lambda (n : T) . b \Rightarrow \mathrm{intro}\ n.\ p }

\inferrule[Symmetry]
  { \Gamma \vdash H \Rightarrow p }
  { \Gamma \vdash \mathtt{eq\_sym}\ H \Rightarrow \mathrm{symmetry}.\ p }

\inferrule[Split]
  { \Gamma \vdash l \Rightarrow p \\ \Gamma \vdash r \Rightarrow q }
  { \Gamma \vdash \mathrm{Constr}(0,\ \wedge)\ l\ r \Rightarrow \mathrm{split} \{ p, q \}.\ }\\

\inferrule[Left]
  { \Gamma \vdash H \Rightarrow p }
  { \Gamma \vdash \mathrm{Constr}(0,\ \vee)\ H \Rightarrow \mathrm{left}.\ p }

\inferrule[Right]
  { \Gamma \vdash H \Rightarrow p }
  { \Gamma \vdash \mathrm{Constr}(1,\ \vee)\ H \Rightarrow \mathrm{right}.\ p }

\inferrule[Rewrite]
  { \Gamma \vdash H_1 : x = y \\ \Gamma \vdash H_2 \Rightarrow p }
  { \Gamma \vdash \mathrm{Elim}(H_1,\ P) \{ x,\ H_2,\ y \} \Rightarrow \mathrm{symmetry}.\ \mathrm{rewrite}\ P\ H_1.\ p }\\

\inferrule[Induction]
  { \Gamma \vdash \vec{f} \Rightarrow \vec{p} }
  { \Gamma \vdash \mathrm{Elim}(t,\ P)\ \vec{f} \Rightarrow \mathrm{induction}\ P\ t\ \vec{p} }

\inferrule[Apply]
  { \Gamma \vdash t \Rightarrow p }
  { \Gamma \vdash f t \Rightarrow \mathrm{apply}\ f.\ p }

\inferrule[Base]
  { \\ }
  { \Gamma \vdash t \Rightarrow \mathrm{apply}\ t }
\end{mathpar}
\vspace{-0.4cm}
\caption{Qtac decompiler semantics.}
\label{fig:someantics}
\end{figure*}

\mysubsubsec{Equality}
To ensure coherence with equality (Figure~\ref{fig:spec}, right),
\lstinline{Eta} and \lstinline{Iota} must prove $\eta$ and $\iota$.
That is, \lstinline{Eta} must have the same definitional behavior as the dependent eliminator (\lstinline{elim_eta}),
and must behave like identity (\lstinline{eta_ok}).
Each \lstinline{Iota} must prove and rewrite along the simplification (\textit{refolding}~\cite{boutillier:tel-01054723}) behavior that corresponds to a case of the dependent eliminator (\lstinline{iota_ok}).
This makes it possible for the transformation to
avoid applying \lstinline{section} and \lstinline{retraction}.

\mysubsubsec{Correctness}
With these correctness criteria for a configuration, we get the completeness result (proven in Coq~\href{https://github.com/uwplse/pumpkin-pi/blob/v2.0.0/plugin/coq/playground/arbitrary.v}{\circled{8}}) that every equivalence induces a configuration. % arbitrary.v
We also obtain an algorithm for the soundness result that every configuration induces an equivalence.

The algorithm to prove \lstinline{section} is as follows (\lstinline{retraction} is similar):
replace \lstinline{a} with \lstinline{Eta(A) a} by \lstinline{eta_ok(A)}.
Then, induct using \lstinline{DepElim} over \A.
For each case $i$, the proof obligation is to show that \lstinline{g (f a)} is equal to \lstinline{a},
where \lstinline{a} is \lstinline{DepConstr(A, i)} applied to the non-inductive arguments (by \lstinline{elim_eta(A)}).
Expand the right-hand side using \lstinline{Iota(A, i)}, then expand it again using \lstinline{Iota(B, i)}
(destructing over each \lstinline{eta_ok} to apply the corresponding \lstinline{Iota}).
The result follows by definition of \lstinline{g} and \lstinline{f}, and by reflexivity.

\mysubsubsec{Automatic Configuration}
\toolname implements four search procedures for automatic configuration~\href{https://github.com/uwplse/pumpkin-pi/blob/v2.0.0/plugin/src/automation/lift/liftconfig.ml}{\circled{6}}.
Three of the four procedures are based on the search procedure from 
\textsc{Devoid}~\cite{Ringer2019},
while the remaining procedure instantiates the types \A and \B of a generic configuration that can be defined inside of Coq directly.
%Two use similar algorithms,
%due to space constraints, 
%we do not discuss these in detail.

The algorithm above is essentially what \textbf{Configure} uses to generate functions \lstinline{f} and \lstinline{g} for these configurations~\href{https://github.com/uwplse/pumpkin-pi/blob/v2.0.0/plugin/src/automation/search/search.ml}{\circled{9}}, % search.ml
and also generate proofs \lstinline{section} and \lstinline{retraction} that these functions form an equivalence~\href{https://github.com/uwplse/pumpkin-pi/blob/v2.0.0/plugin/src/automation/search/equivalence.ml}{\circled{10}}. % equivalence.ml
To minimize dependencies, \toolname does not produce proofs of \lstinline{constr_ok} and \lstinline{elim_ok} directly,
as stating these theorems cleanly would require either a special framework~\cite{tabareau2017equivalences}
or a univalent type theory~\cite{univalent2013homotopy}.
If the proof engineer wishes, it is possible to prove these in individual cases~\href{https://github.com/uwplse/pumpkin-pi/blob/v2.0.0/plugin/coq/playground/arbitrary.v}{\circled{8}}, % arbitray.v
but this is not necessary in order to use \toolname. %---they simply need to hold.

\section{Decompiling Proof Terms to Tactics}
\label{sec:decompiler}

\textbf{Transform} produces a proof term,
while the proof engineer typically writes and maintains proof scripts made up of tactics.
We improve usability thanks to the realization that, since Coq's proof term language Gallina is very structured,
we can decompile these Gallina terms to suggested Ltac proof scripts for the proof engineer to maintain.

%\begin{quote}
%\textbf{Insight 3}: The transformed proof terms can then be translated back to tactic scripts.
%\end{quote}

\textbf{Decompile} implements a prototype of this translation~\href{https://github.com/uwplse/coq-plugin-lib/tree/9ef05815c261de9c99b604c6b581ba1c4fbc1a46/src/coq/decompiler/decompiler.ml}{\circled{11}}: % decompiler.ml
it translates a proof term to a suggested proof script that attempts to prove the same theorem the same way.
Note that this problem is not well defined: while there is always a proof script that 
works (applying the proof term with the \lstinline{apply} tactic), the result is often qualitatively unreadable.
This is the baseline behavior to which the decompiler defaults.
The goal of the decompiler is to improve on that baseline as much as possible,
or else suggest a proof script that is close enough to correct that the proof engineer can
manually massage it into something that works and is maintainable.

\textbf{Decompile} achieves this in two passes: The first pass decompiles proof terms to proof scripts that use a predefined set of tactics.
The second pass improves on suggested tactics by simplifying arguments, substituting tacticals, and using
hints like custom tactics and decision procedures.

\mysubsubsec{First Pass: Basic Proof Scripts}
The first pass takes Coq terms and produces tactics in Ltac, the proof script language for Coq.
Ltac can be confusing to reason about, since Ltac tactics can refer to Gallina terms, and the semantics of Ltac depends both on the
semantics of Gallina and on the implementation of proof search procedures written in OCaml.
To give a sense of how the first pass works without the clutter of these details, we start by defining a mini decompiler that 
implements a simplified version of the first pass.
Section~\ref{sec:second} explains how we scale this to the implementation.

The mini decompiler takes CIC$_{\omega}$ terms and produces tactics in a 
mini version of Ltac which we call Qtac.
The syntax for Qtac is in Figure~\ref{fig:ltacsyntax1}.
Qtac includes hypothesis introduction (\lstinline{intro}),
rewriting (\lstinline{rewrite}), symmetry of equality (\lstinline{symmetry}),
application of a term to prove the goal (\lstinline{apply}), induction (\lstinline{induction}),
case splitting of conjunctions (\lstinline{split}),
constructors of disjunctions (\lstinline{left} and \lstinline{right}), and
composition (\lstinline{.}).
Unlike in Ltac, \lstinline{induction} and \lstinline{rewrite} take a motive explicitly (rather than relying on unification),
and \lstinline{apply} creates a new subgoal for each function argument.
%The implementation reasons about Ltac and so does not make these assumptions.

The semantics for the mini decompiler $\Gamma \vdash t \Rightarrow p$ are in Figure~\ref{fig:someantics} (assuming $=$, \lstinline{eq_sym}, $\wedge$, and $\vee$ are defined as in Coq).
As with the real decompiler, the mini decompiler defaults to the proof script
that applies the entire proof term with \lstinline{apply} (\textsc{Base}).
Otherwise, it improves on that behavior by recursing over the proof term and constructing a proof script using a predefined set of tactics.

\iffalse
\begin{figure*}
\begin{minipage}{0.48\textwidth}
\begin{lstlisting}
fun (@\codesimb{(y0 : list A)}@) =>(@\vspace{-0.04cm}@)
  (@\codesima{list_rect}@) _ _  (fun (@\codesima{a l H}@) =>(@\vspace{-0.04cm}@)
    (@\codesimc{eq_ind_r}@) _ (@\codesimd{eq_refl}@) (@\codesimc{(app_nil_r (rev l) (a::[]))}@))(@\vspace{-0.04cm}@)
    (@\codesime{eq_refl}@)(@\vspace{-0.04cm}@)
    (@\codesima{y0}@)(@\vspace{-0.04cm}@)
\end{lstlisting}
\end{minipage}
\begin{minipage}{0.48\textwidth}
\begin{lstlisting}
(@\vspace{-0.14cm}@)
- (@\codesimb{intro y0.}@) (@\codesima{induction y0 as [a l H|].}@)(@\vspace{-0.04cm}@)
  + (@\codesimc{simpl. rewrite app_nil_r.}@) (@\codesimd{auto.}@)(@\vspace{-0.04cm}@)
  + (@\codesime{auto.}@)(@\vspace{-0.04cm}@)
(@\vspace{-0.14cm}@)
\end{lstlisting}
\end{minipage}
\vspace{-0.3cm}
\caption{Proof term (left) and decompiled proof script (right) for the base case of 
\lstinline{rev_app_distr} (Section~\ref{sec:overview}),  with corresponding terms and tactics 
grouped by color and number.}
\label{fig:rainbow}
\end{figure*}
\fi

For the mini decompiler, this is straightforward: Lambda terms become introduction (\textsc{Intro}).
Applications of \lstinline{eq_sym} become symmetry of equality (\textsc{Symmetry}).
Constructors of conjunction and disjunction map to the respective tactics (\textsc{Split}, \textsc{Left}, and \textsc{Right}).
Applications of equality eliminators compose symmetry (to orient the rewrite direction) with rewrites (\textsc{Rewrite}),
and all other applications of eliminators become induction (\textsc{Induction}).
The remaining applications become apply tactics (\textsc{Apply}).
In all cases, the decompiler recurses, breaking into cases, until only the \textsc{Base}
case holds. % at which point we are done.

While the mini decompiler is very simple, only a few small changes are needed
to move this to Coq.
%The result can already handle some of the example proofs \toolname has produced.
The generated proof term of \lstinline{rev_app_distr} from Section~\ref{sec:overview},
for example, consists only of induction, rewriting, simplification, and reflexivity (solved by \lstinline{auto}).
Figure~\ref{fig:rainbow} shows the proof term for the base case of \lstinline{rev_app_distr} 
alongside the proof script that \toolname suggests.
This script is fairly low-level and close to the proof term, but it is already something that the proof engineer
can step through to understand, modify, and maintain.
There are few differences from the mini decompiler needed to produce this,
for example handling of rewrites in both directions (\lstinline{eq_ind_r} as opposed to \lstinline{eq_ind}),
simplifying rewrites,
and turning applications of \lstinline{eq_refl} into \lstinline{reflexivity} or \lstinline{auto}.

\mysubsubsec{Second Pass: Better Proof Scripts}
The implementation of \textbf{Decompile} first runs something similar to the mini decompiler, then modifies the suggested tactics to produce a more natural proof script~\href{https://github.com/uwplse/coq-plugin-lib/tree/9ef05815c261de9c99b604c6b581ba1c4fbc1a46/src/coq/decompiler/decompiler.ml}{\circled{11}}. % decompiler.ml
For example, it cancels out sequences of \lstinline{intros} and \lstinline{revert},
inserts semicolons, and removes extra arguments to \lstinline{apply} and \lstinline{rewrite}. %, ensuring the result still holds. % TODO rewrite
It can also take tactics from the proof engineer (like part of the old proof script) as hints,
then iteratively replace tactics with those hints, checking for correctness.
This makes it possible for suggested scripts to include custom tactics and decision procedures.
%We omit the details due to space constraints. TODO add back for PLDI?
%Further improvements could come from preserving comments and indentation, or automatically using information from the old 
%version of the proof script rather than asking for it explicitly.

%In fact, since \toolname uses an existing command to translate pattern matching and fixpoints to eliminators,
%\textit{all} of the proof terms that \toolname produces will use induction and rewriting instead.
%Because we have control over output terms, even a mini decompiler gets us pretty far.

% TODO add any new things RanDair implements, like exists

\section{Implementation}
\label{sec:impl}

The transformation and mini decompiler abstract many of the challenges
of building a tool for proof engineers. % many details needed to build a tool that reaches proof engineers.
%and the mini decompiler abstracts a lot of the details that make Ltac so useful to proof engineers---and so painful to 
%reason about automatically.
This section describes how we solved some of these challenges.
%for both the transformation (Section~\ref{sec:implementation}) and the decompiler (Section~\ref{sec:second}).
%Section~\ref{sec:discussion} describes some remaining challenges and our plans to address them. % in the future.

\subsection{Implementing the Transformation}
\label{sec:implementation}

\mysubsubsec{Termination}
When a subterm unifies with a configuration term, this suggests that \toolname \textit{can}
transform the subterm, but it does not necessarily mean that it \textit{should}.
In some cases, doing so would result in nontermination.
For example, if \B is a refinement of \A, then we can always run \textsc{Equivalence}
over and over again, forever.
%\textsc{Devoid} ruled out this case by simply prohibiting the case where \B refers to \A, but we found it sometimes
%useful to support this case.
We thus include some simple termination checks in our code~\href{https://github.com/uwplse/pumpkin-pi/blob/v2.0.0/plugin/src/automation/lift/liftrules.ml}{\circled{12}}. % liftrules.ml

\mysubsubsec{Intent}
Even when termination is guaranteed, whether to transform a subterm depends on intent.
That is, \toolname automates the case of porting \textit{every} \A to \B,
but proof engineers sometimes wish to port only \textit{some} $A$s to $B$s.
\toolname has some support for this using an interactive workflow~\href{https://github.com/uwplse/pumpkin-pi/blob/v2.0.0/plugin/coq/minimal_records.v}{\circled{13}},
with plans for automatic support in the future. % minimal_records.v, but show this
%We helped the proof engineer do this by interacting with \toolname using a particular workflow.

\mysubsubsec{From CIC$_{\omega}$ to Coq}
The implementation~\href{https://github.com/uwplse/pumpkin-pi/blob/v2.0.0/plugin/src/automation/lift/lift.ml}{\circled{4}} % lift.ml
of the transformation handles language differences to scale from CIC$_{\omega}$ to Coq.
We use the existing \lstinline{Preprocess}~\cite{Ringer2019} command to turn pattern matching and fixpoints into 
eliminators.
We handle refolding of constants in constructors using \lstinline{DepConstr}.

\begin{figure}
\begin{lstlisting}
fun (@\codesimb{(y0 : list A)}@) =>(@\vspace{-0.04cm}@)
  (@\codesima{list_rect}@) _ _  (fun (@\codesima{a l H}@) =>(@\vspace{-0.04cm}@)
    (@\codesimc{eq_ind_r}@) _ (@\codesimd{eq_refl}@) (@\codesimc{(app_nil_r (rev l) (a::[]))}@))(@\vspace{-0.04cm}@)
    (@\codesime{eq_refl}@)(@\vspace{-0.04cm}@)
    (@\codesima{y0}@)(@\vspace{-0.04cm}@)
(@\vspace{-0.04cm}@)
- (@\codesimb{intro y0.}@) (@\codesima{induction y0 as [a l H|].}@)(@\vspace{-0.04cm}@)
  + (@\codesimc{simpl. rewrite app_nil_r.}@) (@\codesimd{auto.}@)(@\vspace{-0.04cm}@)
  + (@\codesime{auto.}@)(@\vspace{-0.04cm}@)
\end{lstlisting}
\vspace{-0.3cm}
\caption{Proof term (top) and decompiled proof script (bottom) for the base case of 
\lstinline{rev_app_distr} (Section~\ref{sec:overview}), with corresponding terms and tactics 
grouped by color \& number.}
\label{fig:rainbow}
\end{figure}

\mysubsubsec{Reaching Real Proof Engineers}
Many of our design decisions in implementing \toolname were informed by our partnership with
an industrial proof engineer (Section~\ref{sec:search}).
For example, the proof engineer rarely had the patience to wait more than ten seconds
for \toolname to port a term,
so we implemented optional aggressive caching, even caching intermediate subterms
encountered while running the transformation~\href{https://github.com/uwplse/pumpkin-pi/blob/v2.0.0/plugin/src/cache/caching.ml}{\circled{14}}. % TODO caching.ml
We also added a cache to tell \toolname not to $\delta$-reduce certain terms~\href{https://github.com/uwplse/pumpkin-pi/blob/v2.0.0/plugin/src/cache/caching.ml}{\circled{14}}.
With these caches, the proof engineer found \toolname efficient enough to use on a code base with tens of thousands of lines of code and proof.

 % caching.ml
%These caches are implemented in \href{https://github.com/uwplse/pumpkin-pi/blob/master/plugin/src/cache/caching.ml}{caching.ml}.
%or recurse into certain modules.
% set certain terms or modules as opaque to \toolname, to prevent unnecessary $\delta$-reduction.

The experiences of proof engineers also inspired new features.
For example, we implemented a search procedure to generate custom eliminators %(\href{https://github.com/uwplse/pumpkin-pi/blob/master/plugin/src/automation/search/smartelim.ml}{smartelim.ml})
to help reason about types like $\Sigma$\lstinline{(l : list T).length l = n}
by reasoning separately about the projections~\href{https://github.com/uwplse/pumpkin-pi/blob/v2.0.0/plugin/src/automation/search/smartelim.ml}{\circled{15}}. %smartelim.ml
We added informative error messages~\href{https://github.com/uwplse/pumpkin-pi/blob/v2.0.0/plugin/src/lib/ornerrors.ml}{\circled{22}} to help the proof engineer distinguish between user errors and bugs. % TODO link to errors
These features helped with workflow integration. % tactic decompiler helped with integration into proof engineering workflows.

\begin{table*}
  \caption{Some changes using \toolname (left to right): class of changes, kind of configuration, examples, whether using \toolname saved development time relative to reference manual repairs (\good\xspace if yes, \ok\xspace if comparable, \bad\xspace if no), and Coq tools we know of that support repair along (Repair) or automatic proof of (Search) the equivalence corresponding to each example. Tools considered are \textsc{Devoid}~\cite{Ringer2019}, the Univalent Parametricity (UP) white-box transformation~\cite{tabareau2019marriage}, and the tool from \citet{magaud2000changing}. \toolname is the only one that suggests tactics.
More nuanced comparisons to these and more are in Section~\ref{sec:related}.}
\vspace{-0.35cm}
\small
  \begin{tabular}{|l|l|l|c|l|l|}
    \hline
    \textbf{Class} & \textbf{Config.} & \textbf{Examples} & \textbf{Sav.} & \textbf{Repair Tools} & \textbf{Search Tools} \\
    \hline
    \multirow[t]{2}{*}{Algebraic Ornaments} & \multirow[t]{2}{*}{Auto} & List to Packed Vector, hs-to-coq \href{https://github.com/uwplse/pumpkin-pi/blob/v2.0.0/plugin/coq/examples/Example.v}{\circled{3}} % Example.v
    & \good & \toolname, \textsc{Devoid}, UP & \toolname, \textsc{Devoid} \\
    & & List to Packed Vector, Std. Library \href{https://github.com/uwplse/pumpkin-pi/blob/v2.0.0/plugin/coq/examples/ListToVect.v}{\circled{16}} % ListToVect.v
    & \good & \toolname, \textsc{Devoid}, UP & \toolname, \textsc{Devoid} \\
    \hline
    Unpack Sigma Types & Auto & Vector of Given Length, hs-to-coq \href{https://github.com/uwplse/pumpkin-pi/blob/v2.0.0/plugin/coq/examples/Example.v}{\circled{3}} % Example.v
    & \good & \toolname, UP & \toolname \\
    \hline
    \multirow[t]{3}{*}{Tuples \& Records} & \multirow[t]{3}{*}{Auto} & Simple Records \href{https://github.com/uwplse/pumpkin-pi/blob/v2.0.0/plugin/coq/minimal_records.v}{\circled{13}} % minimal_records.v 
     & \good & \toolname, UP & \toolname \\
    & & Parameterized Records \href{https://github.com/uwplse/pumpkin-pi/blob/v2.0.0/plugin/coq/more_records.v}{\circled{17}} % more_records.v
    & \good & \toolname, UP & \toolname \\
    & & Industrial Use \href{https://github.com/Ptival/saw-core-coq/tree/dump-wip}{\circled{18}} %(\href{https://github.com/Ptival/saw-core-coq/tree/dump-wip}{saw-core-coq})
    & \good & \toolname, UP & \toolname \\
    \hline
    \multirow[t]{3}{*}{Permute Constructors} & \multirow[t]{3}{*}{Auto} & List, Standard Library \href{https://github.com/uwplse/pumpkin-pi/blob/v2.0.0/plugin/coq/Swap.v}{\circled{1}}
    & \good & \toolname, UP & \toolname \\
     & & Modifying PL, \textsc{REPLica} Benchmark \href{https://github.com/uwplse/pumpkin-pi/blob/v2.0.0/plugin/coq/Swap.v}{\circled{1}} % Swap.v 
     & \ok & \toolname, UP  & \toolname \\
    & & Large Ambiguous Enum \href{https://github.com/uwplse/pumpkin-pi/blob/v2.0.0/plugin/coq/Swap.v}{\circled{1}} % Swap.v
    & \ok & \toolname, UP & \toolname \\
    \hline
    Add new Constructors & Mixed & PL Extension, \textsc{REPLica} Benchmark \href{https://github.com/uwplse/pumpkin-pi/blob/v2.0.0/plugin/coq/playground/add_constr.v}{\circled{19}} % (\href{https://github.com/uwplse/pumpkin-pi/blob/master/plugin/coq/playground/add_constr.v}{add_constr.v})
    & \bad & \toolname & \toolname (partial) \\
    \hline
    Factor Constructors & Manual & External Example \href{https://github.com/uwplse/pumpkin-pi/blob/v2.0.0/plugin/coq/playground/constr_refactor.v}{\circled{2}} % (\href{https://github.com/uwplse/pumpkin-pi/blob/master/plugin/coq/playground/constr_refactor.v}{constr_refactor.v}) 
    & \good & \toolname, UP & None \\
    \hline
    Permute Hypotheses & Manual & External Example \href{https://github.com/uwplse/pumpkin-pi/blob/v2.0.0/plugin/coq/playground/flip.v}{\circled{20}} %(\href{https://github.com/uwplse/pumpkin-pi/blob/master/plugin/coq/playground/flip.v}{flip.v}) 
    & \bad & \toolname, UP & None \\
    \hline
    \multirow[t]{2}{*}{Change Ind. Structure} & \multirow[t]{2}{*}{Manual} & Unary to Binary, Classic Benchmark \href{https://github.com/uwplse/pumpkin-pi/blob/v2.0.0/plugin/coq/nonorn.v}{\circled{5}} %(\href{https://github.com/uwplse/pumpkin-pi/blob/master/plugin/coq/nonorn.v}{nonorn.v})
     & \ok & \toolname, Magaud & None \\
     & & Vector to Fin. Set, External Example \href{https://github.com/uwplse/pumpkin-pi/blob/v2.0.0/plugin/coq/playground/fin.v}{\circled{21}} % (\href{https://github.com/uwplse/pumpkin-pi/blob/master/plugin/coq/playground/fin.v}{fin.v}) 
     & \good & \toolname & None \\
    \hline
  \end{tabular}
\label{fig:changes}
\end{table*}

\subsection{Implementing the Decompiler}
\label{sec:second}

\mysubsubsec{From Qtac to Ltac}
The mini decompiler assumes more predictable versions of \lstinline{rewrite} and \lstinline{induction}
than those in Coq. \textbf{Decompile} includes additional logic to reason about these tactics~\href{https://github.com/uwplse/coq-plugin-lib/blob/9ef05815c261de9c99b604c6b581ba1c4fbc1a46/src/coq/decompiler/decompiler.ml}{\circled{11}}. % decompiler.ml
For example, Qtac assumes that there is only one \lstinline{rewrite} direction. Ltac has two rewrite directions,
and so the decompiler infers the direction from the motive.

Qtac also assumes that both tactics take the inductive motive explicitly,
while in Coq, both tactics infer the motive automatically.
Consequentially, Coq sometimes fails to infer the correct motive.
% infers the wrong motive, % without manipulation of goals and hypotheses,
%or fails to infer a motive at all.
%This is especially common for \lstinline{rewrite}, which is purely syntactic.
To handle induction, the decompiler strategically uses \lstinline{revert} to manipulate the goal
so that Coq can better infer the motive.
To handle rewrites, it uses \lstinline{simpl} to refold the goal before rewriting.
Neither of these approaches is guaranteed to work, so the proof engineer may sometimes need to tweak the suggested proof script appropriately.
Even if we pass Coq's induction principle an explicit motive, Coq still sometimes fails due
to unrepresented assumptions.
Long term, using another tactic like \lstinline{change} or \lstinline{refine} before applying these tactics
may help with cases for which Coq cannot infer the correct motive.

\mysubsubsec{From CIC$_{\omega}$ to Coq}
Scaling the decompiler to Coq introduces let bindings, which are generated by 
tactics like \lstinline{rewrite in}, \lstinline{apply in}, and \lstinline{pose}.
\textbf{Decompile} implements~\href{https://github.com/uwplse/coq-plugin-lib/blob/9ef05815c261de9c99b604c6b581ba1c4fbc1a46/src/coq/decompiler/decompiler.ml}{\circled{11}} % decompiler.ml
support for \lstinline{rewrite in} and \lstinline{apply in} similarly to how it supports
\lstinline{rewrite} and \lstinline{apply}, except that it ensures that the unmanipulated hypothesis does not occur in the body of the let expression,
it swaps the direction of the rewrite, and it recurses into any generated subgoals.
In all other cases, it uses \lstinline{pose}, a catch-all for let bindings.

\mysubsubsec{Forfeiting Soundness}
While there is a way to always produce a correct proof script,
\textbf{Decompile} deliberately forfeits soundness to suggest more useful tactics.
For example, it may suggest the \lstinline{induction} tactic, but leave the step of motive inference to the proof engineer.
We have found these suggested tactics easier to work with (Section~\ref{sec:search}).
Note that in the case the suggested proof script is not quite correct,
it is still possible to use the generated proof term directly.

\mysubsubsec{Pretty Printing}
After decompiling proof terms, \textbf{Decompile} pretty prints the result~\href{https://github.com/uwplse/coq-plugin-lib/blob/9ef05815c261de9c99b604c6b581ba1c4fbc1a46/src/coq/decompiler/decompiler.ml}{\circled{11}}.
Like the mini decompiler, \textbf{Decompile} represents its output using a predefined grammar of Ltac tactics,
albeit one that is larger than Qtac, and that also includes tacticals.
It maintains the recursive proof structure for formatting. %, then uses that to print proofs of subgoals using bullet points.
%It displays the resulting proof script to the proof engineer, who can modify it as needed.
%It includes scripts that automate the process of printing all of these tactic proofs to a Coq file,
%in case the proof engineer does not want an interactive workflow.
\toolname keeps all output terms from \textbf{Transform} in the Coq environment in case the decompiler does not succeed.
Once the proof engineer has the new proof, she can remove the old one.

\section{Case Studies: \textsc{Pumpkin P}i Eight Ways}
\label{sec:search}

This section summarizes eight case studies using \toolname,
corresponding to the eight rows in Table~\ref{fig:changes}.
%For each case study, we explain the configuration used, walk through an example, and describe lessons learned.
These case studies highlight \toolname's flexibility in handling diverse scenarios,
%including in one case for an industrial user with unanticipated workflow.
the success of automatic configuration for better workflow integration, % for supported changes,
the preliminary success of the prototype decompiler,
and clear paths to better serving proof engineers.
Detailed walkthroughs are in the code.

\mysubsubsec{Algebraic Ornaments: Lists to Packed Vectors}
The transformation in \toolname is a generalization of the transformation from \textsc{Devoid}.
\textsc{Devoid} supported proof reuse across \textit{algebraic ornaments}, which describe relations
between two inductive types, where one type is the other indexed by a fold~\cite{mcbride}.
A standard example is the relation between a list and a
length-indexed vector (Figure~\ref{fig:listtovect}).

\toolname implements a search procedure for automatic configuration of algebraic ornaments.
%The search procedure moves the work specific to algebraic ornaments into the \toolname configuration.
The result is all functionality from \textsc{Devoid}, plus tactic suggestions.
%which allows it to implement the  functionality from \textsc{Devoid}.
In file~\href{https://github.com/uwplse/pumpkin-pi/blob/v2.0.0/plugin/coq/examples/Example.v}{\circled{3}}, we used this to port
functions and a proof from lists to vectors of \textit{some} length, since \lstinline{list T} $\simeq$ \lstinline{packed_vect T}.
The decompiler helped us write proofs in the order of hours that we had found
too hard to write by hand,
though the suggested tactics did need massaging.
% as the decompiler struggled with motive inference
%for induction with dependent types.
%Additional effort is needed to improve tactic suggestions for dependent types.

\begin{figure*}
\begin{minipage}{0.48\textwidth}
\begin{lstlisting}[firstline=1, lastline=9]
Inductive Term : Set :=(@\vspace{-0.04cm}@)
| Var : Identifier $\rightarrow$ Term(@\vspace{-0.04cm}@)
| (@\codediff{Int}@) : Z $\rightarrow$ Term(@\vspace{-0.04cm}@)
| (@\codediff{Eq}@) : Term $\rightarrow$ Term $\rightarrow$ Term(@\vspace{-0.04cm}@)
| Plus : Term $\rightarrow$ Term $\rightarrow$ Term(@\vspace{-0.04cm}@)
| Times : Term $\rightarrow$ Term $\rightarrow$ Term(@\vspace{-0.04cm}@)
| Minus : Term $\rightarrow$ Term $\rightarrow$ Term(@\vspace{-0.04cm}@)
| Choose : Identifier $\rightarrow$ Term $\rightarrow$ Term.(@\vspace{-0.04cm}@)
(@\vspace{-0.18cm}@)
\end{lstlisting}
\end{minipage}
\hfill
\begin{minipage}{0.48\textwidth}
\begin{lstlisting}[firstline=10, lastline=18]
Inductive Term : Set :=(@\vspace{-0.04cm}@)
| Var : Identifier $\rightarrow$ Term(@\vspace{-0.04cm}@)
| (@\codediff{Bool}@) : Identifier $\rightarrow$ Term(@\vspace{-0.04cm}@)
| (@\codediff{Eq}@) : Term $\rightarrow$ Term $\rightarrow$ Term(@\vspace{-0.04cm}@)
| (@\codediff{Int}@) : Z $\rightarrow$ Term(@\vspace{-0.04cm}@)
| Plus : Term $\rightarrow$ Term $\rightarrow$ Term(@\vspace{-0.04cm}@)
| Times : Term $\rightarrow$ Term $\rightarrow$ Term(@\vspace{-0.04cm}@)
| Minus : Term $\rightarrow$ Term $\rightarrow$ Term(@\vspace{-0.04cm}@)
| Choose : Identifier $\rightarrow$ Term $\rightarrow$ Term.(@\vspace{-0.04cm}@)
\end{lstlisting}
\end{minipage}
\vspace{-0.3cm}
\caption{A simple language (left) and the same language with two swapped constructors and an added constructor (right).}
\vspace{0.1cm}
\label{fig:replica}
\end{figure*}

\mysubsubsec{Unpack Sigma Types: Vectors of Particular Lengths}
%The previous case study showed how to get between lists and vectors of \textit{some} length.
In the same file~\href{https://github.com/uwplse/pumpkin-pi/blob/v2.0.0/plugin/coq/examples/Example.v}{\circled{3}}, we then ported functions and proofs to vectors of a \textit{particular} length, 
like \lstinline{vector T n}.
\textsc{Devoid} had left this step to the proof engineer.
We supported this in \toolname by chaining the previous change
with an automatic configuration for unpacking sigma types.
%To support this, we used one additional automatic configuration for unpacking sigma types.
%This configuration corresponds to the equivalence between sigma types at a particular projection
%and the same type escaping the sigma type. %, in our example $\Sigma$\lstinline{(s : packed_vect T).}$\pi_l$\lstinline{ s = n }$\simeq$\lstinline
%{ vector T n}.
By composition, this transported proofs across the equivalence from Section~\ref{sec:key1}.

Two tricks helped with better workflow integration for this change:
1) have the search procedure view \lstinline{vector T n} as 
$\Sigma$\lstinline{(v : vector T m).n = m} for some \lstinline{m},
then let \toolname instantiate those equalities via unification heuristics, %before transforming,
and 2) generate a custom eliminator for combining
list terms with length invariants.
%This gave us a proof of this lemma (with \lstinline{zip_with} and \lstinline{zip} operating over lists at given lengths):
The resulting workflow works not just for lists and vectors, but for any algebraic ornament,
automating manual effort from \textsc{Devoid}.
The suggested tactics were helpful for writing proofs in the order of hours
that we had struggled with manually over the course of days, but only after massaging.
More effort is needed to improve tactic suggestions for dependent types.

\mysubsubsec{Tuples \& Records: Industrial Use}
An industrial proof engineer at the company \company has been using \toolname in proving
correct an implementation of the TLS handshake protocol.
%While this is ongoing work, thus far,
%\toolname has helped \company integrate Coq with their existing verification workflow.
\company had been using a custom solver-aided verification language to prove correct C programs,
but had found that at times, the constraint solvers got stuck.
%and they could not progress on proofs about those programs.
They had built a compiler that translates their language into Coq's specification language Gallina,
that way proof engineers could finish stuck proofs interactively using Coq.
However, due to language differences, they had found the generated Gallina programs and specifications difficult to work with.

The proof engineer used \toolname to port the automatically generated functions and specifications to more
human-readable functions and specifications, wrote Coq proofs about those functions and specifications, then
used \toolname to port those proofs back to
proofs about the original functions and specifications.
%This workflow has allowed for industrial integration with Coq and has helped the proof engineer write functions and proofs
%that would have otherwise been difficult.
So far, they have used at least three automatic configurations,
but they most often used an automatic configuration for porting compiler-produced anonymous tuples
to named records, as in file~\href{https://github.com/Ptival/saw-core-coq/tree/dump-wip}{\circled{18}}. % TODO aux material %\footnote{\url{https://github.com/Ptival/saw-core-coq/tree/dump-wip}} TODO note in guide it got a bit abandoned because proof engineer got stuck in france...
%The proof engineer was able to use \toolname to integrate Coq into an existing proof engineering
%workflow using solver-aided tools at \company.
The workflow was a bit nonstandard,
so there was little need for tactic suggestions.
The proof engineer reported an initial time investment learning how to use \toolname,
followed by later returns.
%In the initial days, we worked closely with the proof engineer;
%later, the proof engineer worked independently and reached out occasionally.
%The proof engineer was able to work independently, but found two challenges with workflow integration:
%1) they sometimes could not distinguish between user errors and bugs,
%and 2) they waited only about ten seconds at most for \toolname to return.
%Both informed improvements to \toolname, like better error messages, caching,
%and a way to tell \toolname not to $\delta$-reduce certain terms.

\mysubsubsec{Permute Constructors: Modifying a Language}
The swapping example from Section~\ref{sec:overview} was inspired by benchmarks 
from the \textsc{Replica} user study of proof engineers~\cite{replica}.
A change from one of the benchmarks is in Figure~\ref{fig:replica}.
The proof engineer had a simple language represented by an inductive type \lstinline{Term},
as well as some definitions and proofs about the language.
The proof engineer swapped two constructors in the language,
and added a new constructor \lstinline{Bool}.

This case study and the next case study break this change into two parts.
In the first part, we used \toolname with automatic configuration to repair functions and proofs about the language
after swapping the constructors~\href{https://github.com/uwplse/pumpkin-pi/blob/v2.0.0/plugin/coq/Swap.v}{\circled{1}}.
%We also succeeded at more difficult variants of this,
%like swapping constructors with the same type, renaming all of the constructors,
%permuting more than two constructors,
%or permuting and renaming constructors at the same time.
With a bit of human guidance to choose the permutation from a list of suggestions,
\toolname repaired everything,
though the original tactics would have also worked,
so there was not a difference in development time.
%This is a property of the particular proofs that we had access to;
%As Section~\ref{sec:overview}
%and more advanced variants of this benchmark show~\circled{1}, even for simple changes, this is not always true.
% TODO what happens with the tactic decompiler for these?
%The entire \lstinline{Swap.v} file, which includes swapping constructors of every function in the \lstinline{list} module and
%its dependencies, four variants of the \textsc{Replica} swapping change,
%and testing a large and ambiguous permutation of a 30 constructor \lstinline{Enum},
%took \toolname less than 90 seconds total. % TODO specs, reproduction
%Each variant of the \textsc{Replica} swapping change took \toolname less than 5 seconds, % TODO specs, reproduction
%and the change adding a constructor took \toolname less than 30 seconds. % TODO specs, reproduction

\mysubsubsec{Add new Constructors: Extending a Language}
We then used \toolname to repair functions 
after adding the new constructor in Figure~\ref{fig:replica}, separating out the proof obligations for the new constructor from the old terms~\href{https://github.com/uwplse/pumpkin-pi/blob/v2.0.0/plugin/coq/playground/add_constr.v}{\circled{19}}.
%The resulting functions were guaranteed to preserve the behavior of the old terms. % TODO do the proofs too
This change combined manual and automatic configuration.
We defined an inductive type \lstinline{Diff} and (using partial automation) a configuration to port the terms across the equivalence \lstinline{Old.Term + Diff} $\simeq$ \lstinline{New.Term}.
This resulted in case explosion, but was formulaic, and pointed to a clear path for automation of this class of changes.
The repaired functions guaranteed preservation of the behavior of the original functions. %over \lstinline{Old.Term} before the change.

Adding constructors was less simple than swapping.
For example, \toolname did not yet save us time over the proof engineer from the user study;
fully automating the configuration would have helped significantly.
In addition, the repaired terms were (unlike in the swap case) inefficient compared to human-written terms.
For now, they make good regression tests for the human-written terms---in the future,
we hope to automate the discovery of the more efficient terms,
or use the refinement framework CoqEAL~\cite{cohen:hal-01113453}
to get between proofs of the inefficient and efficient terms.

\vspace{0.105cm}
\mysubsubsec{Factor out Constructors: External Example}
The change from Figure~\ref{fig:equivalence2} came at the request of a non-author.
We supported this using a manual configuration that described which constructor to map to \lstinline{true}
and which constructor to map to \lstinline{false}~\href{https://github.com/uwplse/pumpkin-pi/blob/v2.0.0/plugin/coq/playground/constr_refactor.v}{\circled{2}}.
The configuration was very simple for us to write, and the repaired tactics were immediately useful.
The development time savings were on the order of minutes for a small proof development.
Since most of the modest development time went into writing the configuration,
we expect time savings would increase for a larger development.

\vspace{0.105cm}
\mysubsubsec{Permute Hypotheses: External Example}
The change in \href{https://github.com/uwplse/pumpkin-pi/blob/v2.0.0/plugin/coq/playground/flip.v}{\circled{20}} came at the request of a different non-author (a cubical type theory expert),
and shows how to  use \toolname to swap two hypotheses of a type, since \lstinline{T1} $\rightarrow$ \lstinline{T2} $\rightarrow$ \lstinline{T3} $\simeq$
\lstinline{T2} $\rightarrow$ \lstinline{T1} $\rightarrow$ \lstinline{T3}.
This configuration was manual.
Since neither type was inductive, this change used the generic construction for any equivalence.
%instantiated to this particular equivalence.
This worked well, but necessitated some manual annotation due to the lack of custom unification heuristics for 
manual configuration, and so did not yet save development time, and likely still would not have had the proof development been larger.
Supporting custom unification heuristics would improve this workflow.

\vspace{0.105cm}
\mysubsubsec{Change Inductive Structure: Unary to Binary}
In \href{https://github.com/uwplse/pumpkin-pi/blob/v2.0.0/plugin/coq/nonorn.v}{\circled{5}}, we used \toolname to support a classic example of changing inductive structure:
updating unary to binary numbers,
as in Figure~\ref{fig:nattobin}.
% TODO fin and vect? mention somewhere?
Binary numbers allow for a fast addition function, found in the Coq standard library.
In the style of \citet{magaud2000changing}, we used \toolname to derive a slow binary
addition function that does not refer to \lstinline{nat},
and to port proofs from unary to slow binary addition.
We then showed that the ported theorems hold over fast binary addition.

%The implementation of this is in \lstinline{nonorn.v}.
The configuration for \lstinline{N} used definitions from the Coq standard library
for \lstinline{DepConstr} and \lstinline{DepElim} that had the desired behavior with no changes.
\lstinline{Iota} over the successor case was a rewrite by a lemma
from the standard library that reduced the successor case of the eliminator that we used for \lstinline{DepElim}:

\begin{lstlisting}
N.peano_rect_succ : $\forall$ P pO pS n,(@\vspace{-0.04cm}@)
  N.peano_rect P pO pS (N.succ n) =(@\vspace{-0.04cm}@)
  pS n (N.peano_rect P pO pS n).(@\vspace{-0.05cm}@)
\end{lstlisting}
%
%\begin{lstlisting}
%Lemma iota_1 :(@\vspace{-0.04cm}@)
%  $\forall$ P pO pS n (Q : P (dep_constr_1 n) $\rightarrow$ Type),(@\vspace{-0.04cm}@)
%     Q (pS n (dep_elim P pO pS n)) $\rightarrow$(@\vspace{-0.04cm}@)
%     Q (dep_elim P pO pS (dep_constr_1 n)).(@\vspace{-0.04cm}@)
%Proof.(@\vspace{-0.04cm}@)
%  intros. unfold dep_elim, dep_constr_1. rewrite N.peano_rect_succ. auto.(@\vspace{-0.04cm}@)
%Defined.
%\end{lstlisting}
The need for nontrivial \lstinline{Iota} comes from the fact that \lstinline{N} and \lstinline{nat}
have different inductive structures.
By writing a manual configuration with this \lstinline{Iota}, it was possible for us to implement this transformation 
that had been its own tool.

While porting addition from \lstinline{nat} to \lstinline{N} was automatic after configuring \toolname,
porting proofs about addition took more work.
Due to the lack of unification heuristics for manual configuration,
we had to annotate the proof term to tell \toolname that implicit casts in the inductive cases of proofs were applications of \lstinline{Iota}
over \lstinline{nat}.
These annotations were formulaic, but tricky to write.
Unification heuristics would go a long way toward improving the workflow. % for this use case.

After annotating, we obtained automatically repaired proofs about slow binary addition,
which we found simple to port to fast binary addition.
We hope to automate this last step in the future using CoqEAL. %~\cite{cohen:hal-01113453}.
Repaired tactics were partially useful, but failed to understand custom eliminators like \lstinline{N.peano_rect}, and to generate useful
tactics for applications of \lstinline{Iota}; both of these are clear paths to more useful tactics.
The development time for this proof with \toolname was comparable to reference manual repairs by external proof engineers.
Custom unification heuristics would help bring returns on investment for experts in this use case.

\section{Related Work}
\label{sec:related}

%We discuss related work in proof repair, proof refactoring, proof reuse, and proof design.
%More can be found in a recent survey of proof engineering~\cite{PGL-045}.

\mysubsubsec{Proof Repair}
%\toolname is not the first proof repair tool.
The search procedures in \textbf{Configure} are based partly on ideas from the original \textsc{Pumpkin Patch} prototype~\cite{pumpkinpatch}.
%which includes similar search procedures for discovering patches to fix broken proofs.
The \textsc{Pumpkin Patch} prototype did not apply the patches that it finds,
handle changes in structure, or include support for tactics beyond the use of hints.
%\toolname addresses these limitations.

Proof repair can be viewed as a form of \textit{program repair}~\cite{Monperrus:2018:ASR:3177787.3105906, Gazzola:2018:ASR:3180155.3182526}
for proof assistants.
Proof assistants like Coq are a good fit for program repair: A recent paper~\cite{Qi:2015:APP:2771783.2771791} 
recommends that program repair tools draw on extra information
such as specifications or example patches. In Coq, specifications and examples 
are rich and widely available: specifications thanks to dependent types,
and examples thanks to constructivism.

\mysubsubsec{Proof Refactoring}
%Proof repair is related to proof refactoring~\cite{WhitesidePhD}. 
%and a number of the changes that \toolname supports can be viewed as refactorings.
The proof refactoring tool Levity~\cite{Bourke12} for Isabelle/HOL has seen large-scale industrial use.
Levity focuses on a different task: moving lemmas.
Chick~\cite{robert2018} and RefactorAgda~\cite{wibergh2019} are proof refactoring tools
in a Gallina-like language and in Agda, respectively.
% that also support a few changes that can be viewed as repairs~\cite{PGL-045}.
%Chick operates over a Gallina-like language, while RefactorAgda is implemented in Agda.
These tools support primarily syntactic changes and do not have tactic support.
% changes these tools support are still primarily syntactic,
%and neither of these tools have tactic support.

A few proof refactoring tools operate directly over tactics:
POLAR~\cite{Dietrich2013} refactors proof scripts in languages based on Isabelle/Isar~\cite{Wenzel2007isar},
CoqPIE~\cite{Roe2016} is an IDE with support for simple refactorings of Ltac scripts, and
Tactician~\cite{adams2015} is a refactoring tool for switching between tactics and tacticals.
This approach is not tractable for more complex changes~\cite{robert2018}.

\vspace{0.281cm}
\mysubsubsec{Proof Reuse}
%Proof repair is proof reuse with the additional constraint that one specification ceases to exist.
A few proof reuse tools work by proof term transformation and so can be used for repair.
\citet{Johnsen2004} describes a transformation that generalizes theorems in Isabelle/HOL.
\toolname generalizes the transformation from \textsc{Devoid}~\cite{Ringer2019},
which transformed proofs along algebraic ornaments~\cite{mcbride}.
\citet{magaud2000changing} implement a proof term transformation between
unary and binary numbers. 
Both of these fit into \toolname configurations,
and none suggests tactics in Coq like \toolname does.
The expansion algorithm from \citet{magaud2000changing} may help guide the design
of unification heuristics in \toolname.

The widely used Transfer~\cite{Huffman2013} package supports proof reuse in Isabelle/HOL.
Transfer works by combining a set of extensible transfer rules with a type inference algorithm.
Transfer is not yet suitable for repair, as it necessitates maintaining references to both datatypes.
%In addition, the proof assistant Isabelle/HOL that Transfer works for lacks both dependent types and proof terms.
One possible path toward implementing proof repair in Isabelle/HOL may be to reify proof terms using something like
Isabelle/HOL-Proofs, apply a transformation based on Transfer, and then (as in \toolname) decompile those terms to automation that does not apply Transfer or refer to the old datatype in any way.

CoqEAL~\cite{cohen:hal-01113453} transforms functions across relations in Coq,
and these relations can be more general than \toolname's equivalences.
However, while \toolname supports both functions and proofs, CoqEAL supports only simple functions
due to the problem that \lstinline{Iota} addresses.
CoqEAL may be most useful to chain with \toolname to get faster functions.
Both CoqEAL and recent ornaments work~\cite{williamsphd} may help with
better workflow support for changes that do not correspond to equivalences.

The \toolname transformation implements transport.
Transport is realizable as a function given univalence~\cite{univalent2013homotopy}.
UP~\cite{tabareau2017equivalences} approximates it
in Coq, only sometimes relying on functional extensionality.
While powerful, neither approach removes references to the old type. %making them poorly suited for repair.

Recent work~\cite{tabareau2019marriage} extends UP with 
a white-box transformation that may work for repair.
This imposes proof obligations on the proof engineer beyond those imposed by \toolname,
%that establish what is effectively the correctness criteria
%for the configuration in \toolname, while \toolname needs only that it holds metatheoretically.
and it includes neither search procedures for equivalences nor tactic script generation.
It also does not support changes in inductive structure,
instead relying on its original black-box functionality;
\lstinline{Iota} solves this in \toolname. % and is based on lessons learned from reading that article.
The most fruitful progress may come from combining these tools. % to take advantage of the benefits of both.

%Univalent parametricity implements type-directed search, a feature that \toolname does not yet support.
%It achieves this using type classes~\cite{Sozeau2008}; this does not always scale well~\cite{tabareau2019marriage}.
%Both \toolname and univalent parametricity could benefit from implementing type-directed search using e-graphs~\cite{egraph1}.
%Of particular interest are those developed for congruence in Cubical Agda~\cite{egraph6},
%which prove the theorem \lstinline{hcongr_ideal}~\cite{egraph7} necessary to use e-graphs not derivable in CIC$_{\omega}$,
%and which should allow for efficient and elegant automatic transport.

\mysubsubsec{Proof Design}
Much work focuses on designing proofs
to be robust to change, rather than fixing broken proofs.
This can take the form of design principles, like using 
information hiding techniques~\cite{Woos:2016:PCF:2854065.2854081, Klein:2014:CFV:2584468.2560537}
or any of the structures~\cite{Chrzaszcz2003, Sozeau2008, Saibi:PhD} for encoding interfaces in Coq.
%thereby localizing the burden of change to the interface.
Design and repair are complementary: design requires foresight, while repair can occur retroactively.
Repair can help with changes that occur outside of the proof engineer's control,
or with changes that are difficult to protect against even with informed design.

Another approach to this is to use heavy proof automation, for example through
program-specific proof automation~\cite{Chlipala:2013:CPD:2584504}
%implementations of decision procedures~\cite{Pugh1991},
or general-purpose hammers~\cite{Blanchette2016b, Blanchette2013, Kaliszyk2014, Czajka2018}.
The degree to which proof engineers rely on automation varies, as seen in the data from a user study~\cite{replica}.
Automation-heavy proof engineering styles localize the burden of change to the automation,
but can result in terms that are large and slow to type check,
and tactics that can be difficult to debug.
While these approaches are complementary, more work is needed for \toolname to better support 
developments in this style.

\section{Conclusions \& Future Work}
\label{sec:discussion}

We combined search procedures for equivalences, a proof term transformation,
and a proof term to tactic decompiler to build \toolname,
a proof repair tool for changes in datatypes.
The proof term transformation implements transport across equivalences in a way that is suitable for repair
and that does not compromise definitional equality.
The resulting tool is flexible and useful for real proof engineering scenarios.

%\toolname has helped an industrial proof engineer integrate Coq with a company workflow,
%and it has supported benchmarks common in the proof engineering community.

% encountered in scaling up the \toolname proof term 
%transformation (Section~\ref{sec:problems}), and how we believe ideas from the rewrite system and constraint
%solver communities can address those challenges and improve the state of the art in proof engineering (Section~\ref{sec:egraph}).
%Our hope is to inspire research communities to come together and open the door to better tools for proof reuse and repair.

\mysubsubsec{Future Work}
Moving forward, our goal is to make proofs easier to repair and reuse regardless of proof engineering expertise and style.
We want to reach more proof engineers, and we want \toolname to integrate seamlessly with Coq.

Three problems that we encountered scaling up the \toolname transformation were lack of type-directed search,
ad hoc termination checks, and inability for proof engineers to add custom unification heuristics.
We hope to solve these challenges using \textit{e-graphs}~\cite{egraph1},
a data structure %that is used in the constraint solver and rewrite system communities 
for managing equivalences
built with these kinds of problems in mind.
% to implement search procedures,
%remove the burden of ad hoc reasoning about termination,
%and make it simple for anyone to extend a system with new
%rewrite rules---even ones that can call out to external procedures~\cite{egraph5} 
%like our unification heuristics.
E-graphs were recently adapted to express path equality in cubical~\cite{egraph6}; we hope to repurpose this insight.

Beyond that, we believe that the biggest gains will come from continuing to improve the prototype decompiler.
Two helpful features would be preserving indentation and comments, and automatically using information from the original proof script rather than asking for hints.
One promising path toward the latter is to integrate the decompiler with a machine learning tool like TacTok~\cite{10.1145/3428299} to rank tactic hints.
Some improvements could also come from better tactics,
or from a more structured tactic language.
Integration with version control systems or with integrated development environments could also help.
With that, we believe that the future of seamless and powerful proof repair and reuse for all is within reach.
%and we hope you will join us.
%We hope you will join us in bringing it to life.

%% Acknowledgments
\begin{acks}                            %% acks environment is optional
                                        %% contents suppressed with 'anonymous'
  %% Commands \grantsponsor{<sponsorID>}{<name>}{<url>} and
  %% \grantnum[<url>]{<sponsorID>}{<number>} should be used to
  %% acknowledge financial support and will be used by metadata
  %% extraction tools.
% TODO Gaetan for sure
% TODO call out Anders and Conor specially, given depth of feedback
% TODO Jasper
This paper has really been a community effort.
Anders M\"ortberg and Conor McBride gave us \textit{hours} worth of detailed feedback that was instrumental to writing this paper.
Nicolas Tabareau helped us understand the need to port definitional to propositional equalities.
Valentin Robert gave us feedback on usability that informed tool design.
Ga\"{e}tan Gilbert, James Wilcox, and Jasper Hugunin all at some point helped us write Coq proofs;
let this be a record that we owe Ga\"{e}tan a beer, and we owe James boba.

And of course, we thank our shepherd Gerwin Klein, and we thank all of our reviewers.
We got other wonderful feedback on the paper from 
Cyril Cohen, Tej Chajed, Ben Delaware, Jacob Van Geffen, Janno, James Wilcox, Chandrakana Nandi, 
Martin Kellogg, Audrey Seo, James Decker,
Ben Kushigian, John Regehr, and Justus Adam.
The Coq developers have for \textit{years} given us frequent and efficient feedback on plugin APIs for tool implementation.
The programming languages community on Twitter (yes, seriously) has also been essential to this effort.
Especially during a pandemic. 
And we'd like to extend a special thank you to Talia's mentor, Derek Dreyer.

We have also gotten a lot of feedback on future ideas that we are excited to pursue.
Carlo Angiuli has helped us understand some beautiful theory beneath our implementation 
(spoiler: we believe the analogy connecting \lstinline{DepConstr} and \lstinline{DepElim} to constructors and eliminators
has formal meaning in terms of initial algebras), and we are excited to integrate these insights into future papers
and use them to generalize our insights.
Alex Polozov helped us sketch out ideas for future work with the decompiler.
And we got wonderful feedback on e-graph integration for future work from 
Max Willsey, Chandrakana Nandi, Remy Wang, Zach Tatlock, Bas Spitters, Steven Lyubomirsky, Andrew Liu, Mike He, Ben Kushigian, 
Gus Smith, and Bill Zorn.

This material is based upon work supported by the \grantsponsor{GS100000001}{National Science Foundation}{http://dx.doi.org/10.13039/100000001} Graduate Research Fellowship under Grant No.~\grantnum{GS100000001}{DGE-1256082}. Any opinions, findings, and conclusions or recommendations expressed in this material are those of the authors and do not necessarily reflect the views of the National Science Foundation. % TODO PEO?
\end{acks}

%% Bibliography
\balance
\bibliography{paper.bib}

%% Appendix
%\appendix
%\section{Appendix}

%Text of appendix \ldots

\end{document}